\documentclass[twocolumn,english,nobibnotes,nofootinbib]{revtex4-1}
\usepackage[T1]{fontenc}
\usepackage[a4paper]{geometry}
\usepackage{float}
\usepackage{textcomp}
\usepackage{amsmath}
\usepackage{amssymb}
\usepackage{graphicx}
\usepackage{esint}
\usepackage{natbib}
\usepackage{color}
\usepackage{mathtools}

\makeatletter
\@ifundefined{textcolor}{}
{%
 \definecolor{BLACK}{gray}{0}
 \definecolor{WHITE}{gray}{1}
 \definecolor{RED}{rgb}{1,0,0}
 \definecolor{GREEN}{rgb}{0,1,0}
 \definecolor{BLUE}{rgb}{0,0,1}
 \definecolor{CYAN}{cmyk}{1,0,0,0}
 \definecolor{MAGENTA}{cmyk}{0,1,0,0}
 \definecolor{YELLOW}{cmyk}{0,0,1,0}
}

\DeclarePairedDelimiterX\braket[2]{\langle}{\rangle}{#1 \delimsize\vert #2}
\newcommand{\reff}[1]{(\ref{#1})}
\begin{document}

\title{Aharonov-Bohm conductance oscillations and current equilibration in local n-p junctions in graphene}

\author{D. \.Zebrowski}
\affiliation{AGH University of Science and Technology, Faculty of Physics and
Applied Computer Science,\\
 al. Mickiewicza 30, 30-059 Krak\'ow, Poland}

\author{A. Mre\'{n}ca-Kolasi\'{n}ska}
\affiliation{AGH University of Science and Technology, Faculty of Physics and
Applied Computer Science,\\
 al. Mickiewicza 30, 30-059 Krak\'ow, Poland}

\author{B. Szafran}
\affiliation{AGH University of Science and Technology, Faculty of Physics and
Applied Computer Science,\\
 al. Mickiewicza 30, 30-059 Krak\'ow, Poland}
\begin{abstract}
We consider a small $p$-type island defined within $n$-type graphene nanoribbon induced by potential
of a floating electrode. In the quantum Hall conditions the island supports persistent currents 
localized at the $n-p$ junction. When coupled to the graphene edge the island acts as an Aharonov-Bohm
interferometer. We evaluate the electrostatic potential induced by the floating gate 
within the ribbon near the charge neutrality point and consider equilibration of the currents 
at both sides of the junction. The incoherent equilibration is introduced by the virtual probes technique.
We describe the evolution of the coherent Aharonov-Bohm conductance oscillations to the quantum Hall
fractional plateaus due to the current equilibration.
\end{abstract}
\maketitle

\section{Introduction}

Graphene \cite{CastroNeto2009} as a chiral Dirac  conductor with 
  mean free path  \cite{Bolotin2008,Banszerus2016} 
and coherence length \cite{Miao} of up to several micrometers  is an excellent medium for studies
of electron optics \cite{ Chen2016,Rickhaus2013,  Taychatanapat2015,   Lee2015, Liu2017}
  and formation of electron interferometers \cite{Shytov2008,Young2009, Grushina2013,Russo2008,Huefner2010,Smirnov2012,Rahman2013,Smirnov2014,Cabosart2014}.
This gapless material can be doped electrostatically with the external gates setting the Fermi energy above or below the Dirac point. With multiple gates one can induce bipolar junctions within the sample. In the quantum Hall conditions the $n-p$ junctions 
confine unidirectional currents  \cite{Taychatanapat2015,Barbier2012,Davies2012,Williams2011,Carmier2011,Chen2012,Cresti2008,Milovanovic2014,Milovanovic2014b,
Rickhaus2015,Yliu2015,Zarenia2013,Oroszlany2008,SnakiKolacha,Muller1992,WilliamsLow2011,Tovari2016},
which can be understood classically as a result of the Lorentz force acting towards the junction at both its sides. 
For a system with a floating gate -- a tip of the atomic force microscope for instance \cite{Bours2017} -- one can induce a  $n-p$ nanojunction in the center of a graphene ribbon \cite{Mrenca2016}. 
With the coupling of the edge to the  nanojunction currents, 
the system in high magnetic field  behaves like a  quantum ring 
with  Aharonov-Bohm  conductance oscillations  due to the interference of the incident currents with the ones circulating around the nanojunction \cite{Mrenca2016}. 
The induced quantum nanoring  \cite{Mrenca2016} is an alternative to the ones formed by etching 
\cite{Russo2008,Huefner2010,Smirnov2012,Rahman2013,Cabosart2014,Smirnov2014}. The etched quantum rings in graphene were extensively studied by theory \cite{Recher2007,Hewageegana2008,Abergel2008,Jackiw2009,Schelter2010,Wurm2010,Wu2010,Downing2011,Faria2013,Costa2014,Rakyta2014}. 

In mesoscopic systems the electron and hole currents co-propagate along the bipolar junction for a sufficiently long distance to activate an incoherent equilibration process \cite{Abanin2007, Williams2007, Ozyilmaz2007, gasse2016}.
 The equilibration divides the currents equally between all the available conducting channels at the physical edge of graphene and the bipolar junction
\cite{Alphenaar1990}. The  process results in
appearance  of fractional plateaus of quantum Hall conductance  \cite{Abanin2007, Williams2007, Ozyilmaz2007, gasse2016, famet}.
The dephasing in the junction currents is particularily effective in the low-mobility samples. For
graphene embedded in van der Waals heterostructures \cite{GeimGrigo}, the
mobility is much higher and the 
junction currents preserve at least partially their phase  \cite{Weie2017,Makk}.
In these conditions n-p-n junctions can be used as Mach-Zehnder interferometers \cite{Weie2017,Morikawa} in the quantum Hall regime. Aharonov-Bohm conductance oscillations for the 
confined loops of currents in these systems have been observed \cite{Weie2017,Morikawa,Makk}.

The purpose of the present paper is to provide a numerical model based on a direct solution of the quantum scattering problem in which both the coherent condcuctance oscillations \cite{Mrenca2016,Morikawa,Weie2017,Makk} due 
to the circulation of the confined currents and the equilibration processes \cite{Abanin2007, Williams2007, Ozyilmaz2007, gasse2016} coexist. 
We investigate the evolution of the conductance maps as functions of the external magnetic field and the Fermi energy from the coherent interference pattern to the fractional plateaus of conductance characteristic to the equilibrated transport.  
Contrary to Ref. \onlinecite{Mrenca2016} where fully phase-coherent transport is investigated, the present model accounts for dephasing with  the B\"uttiker virtual probes \cite{Buttiker1988}. The probes in the present approach shift with the external voltages to follow the position of the n-p junction. Equilibration scenarios for spin-dependent Fermi levels \cite{Weie2017} are discussed. Our approach for evaluation of the plateau structure is deterministic in contrast to the statistical techniques averaging the conductance over many configurations of rough edge termination \cite{Myoung2017}, and random on-site potential on the interface without \cite{Long2008} or with the virtual probes attached \cite{Chen2011}.

The current confinement and equilibration appear along the n-p junctions defined electrostatically in the gapless graphene and the exact
spatial profile determines the visibility and the period of the Aharonov-Bohm oscillations.  Previously, an isotropic model potential for the tip potential was used in Ref. \onlinecite{Mrenca2016} for a proof-of-principle calculation. In fact the electrostatic potential landscape within graphene is
a result of the screening of the long range Coulomb potential of the charged tip. Since in the quantum Hall conditions the edges 
of the sample form waveguides to the currents, the finite size of the sample needs to be included in the modeling. For that reason a graphene nanoribbons are considered below. The presence of the ribbon edges makes the screening of the potential anisotropic which calls for evaluation of the effective potential that we provide in this work using a Schr\"odinger-Poisson scheme. We find that the screening is more effective for metallic than for  semiconducting ribbons. The model allows for evaluation of the potential profile for an arbitrary top gate. We also compare the results obtained for the tip  the rectangular n-p-n junctions  as applied in Refs. \cite{Weie2017,Makk,Morikawa}.

\section{Theory}
 
\subsection{Modeling the potential profile}
\begin{figure}[htbp]
\includegraphics[width=0.25\textwidth]{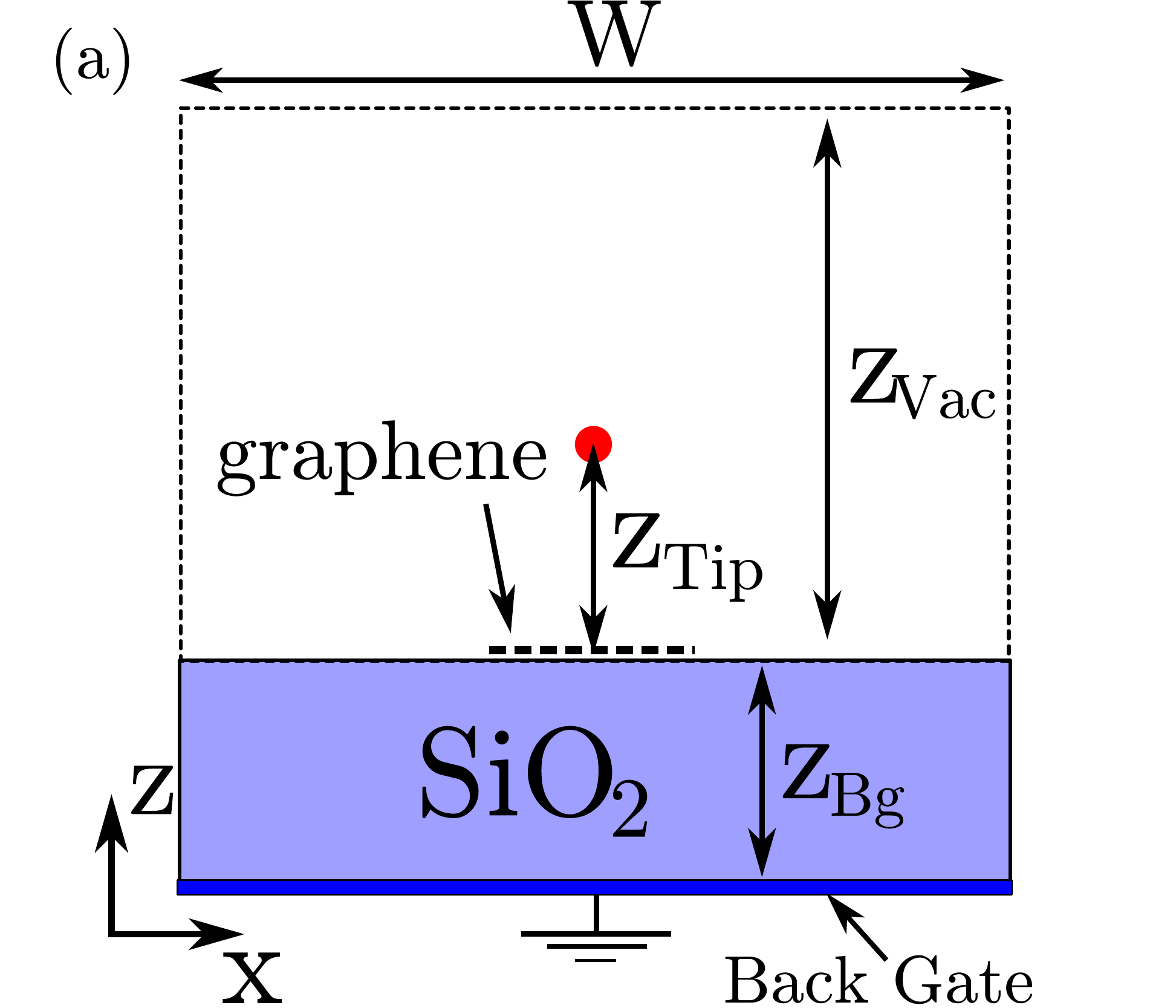}\includegraphics[width=0.25 \textwidth]{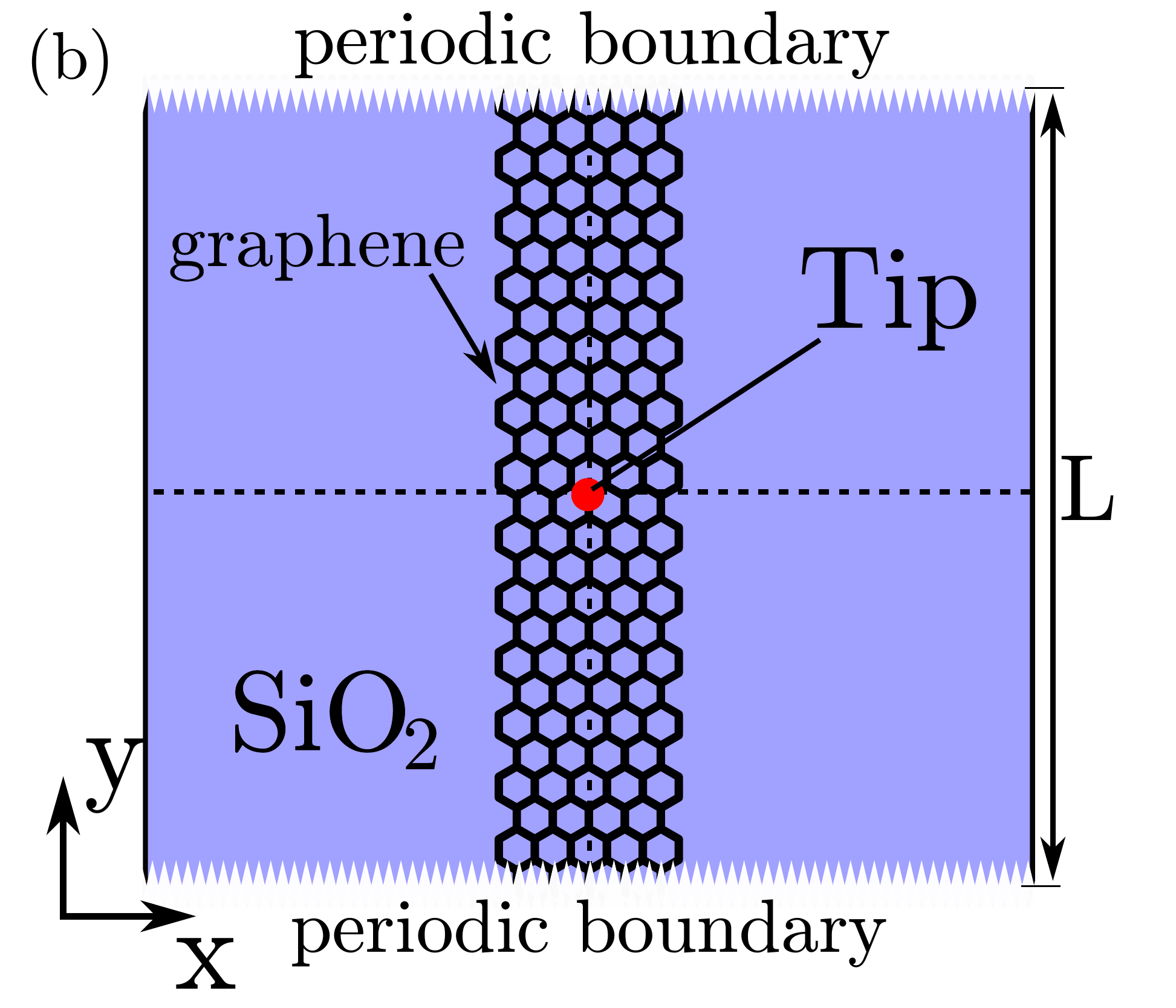}
\newline
\newline
\newline
\includegraphics[width=0.25\textwidth]{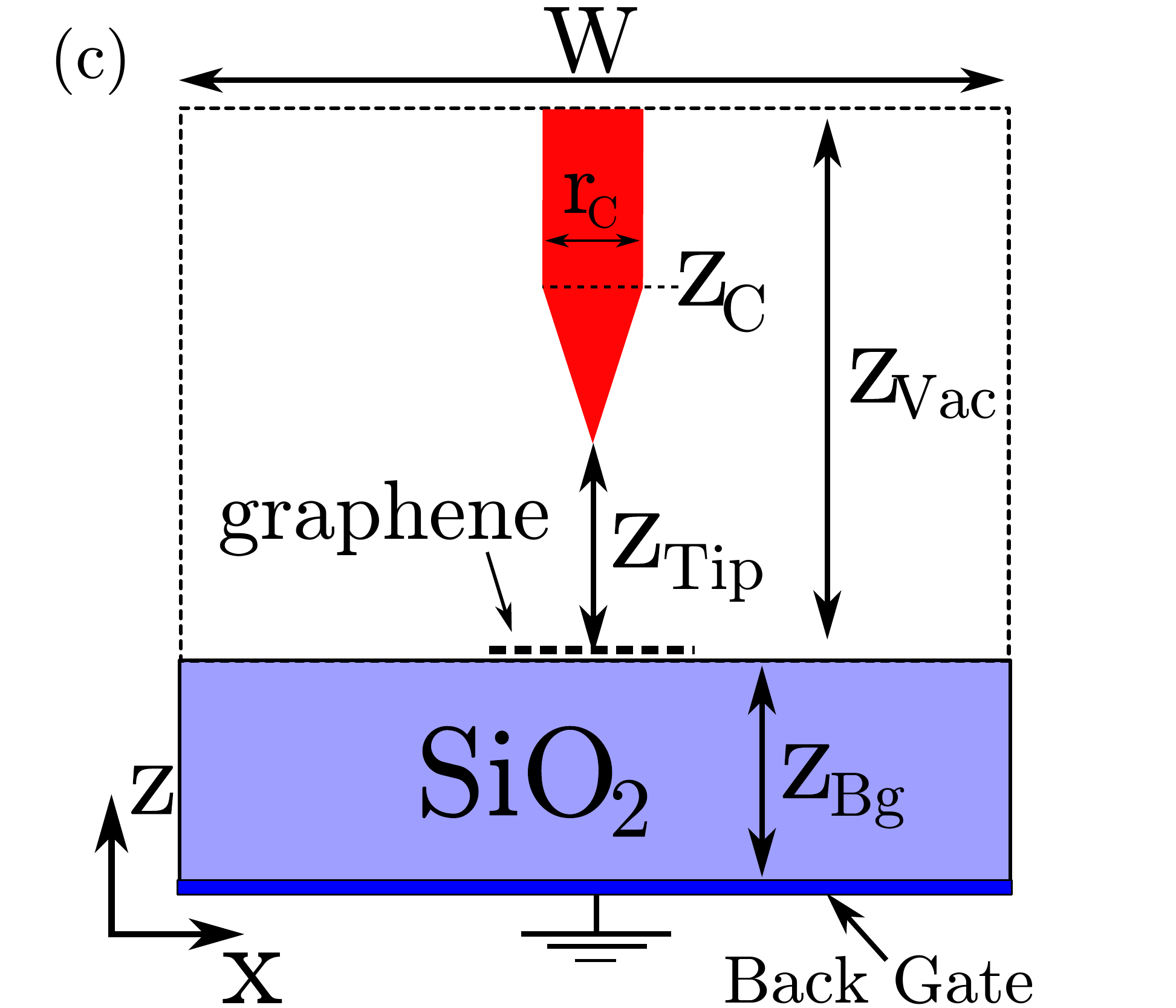}\includegraphics[width=0.25 \textwidth]{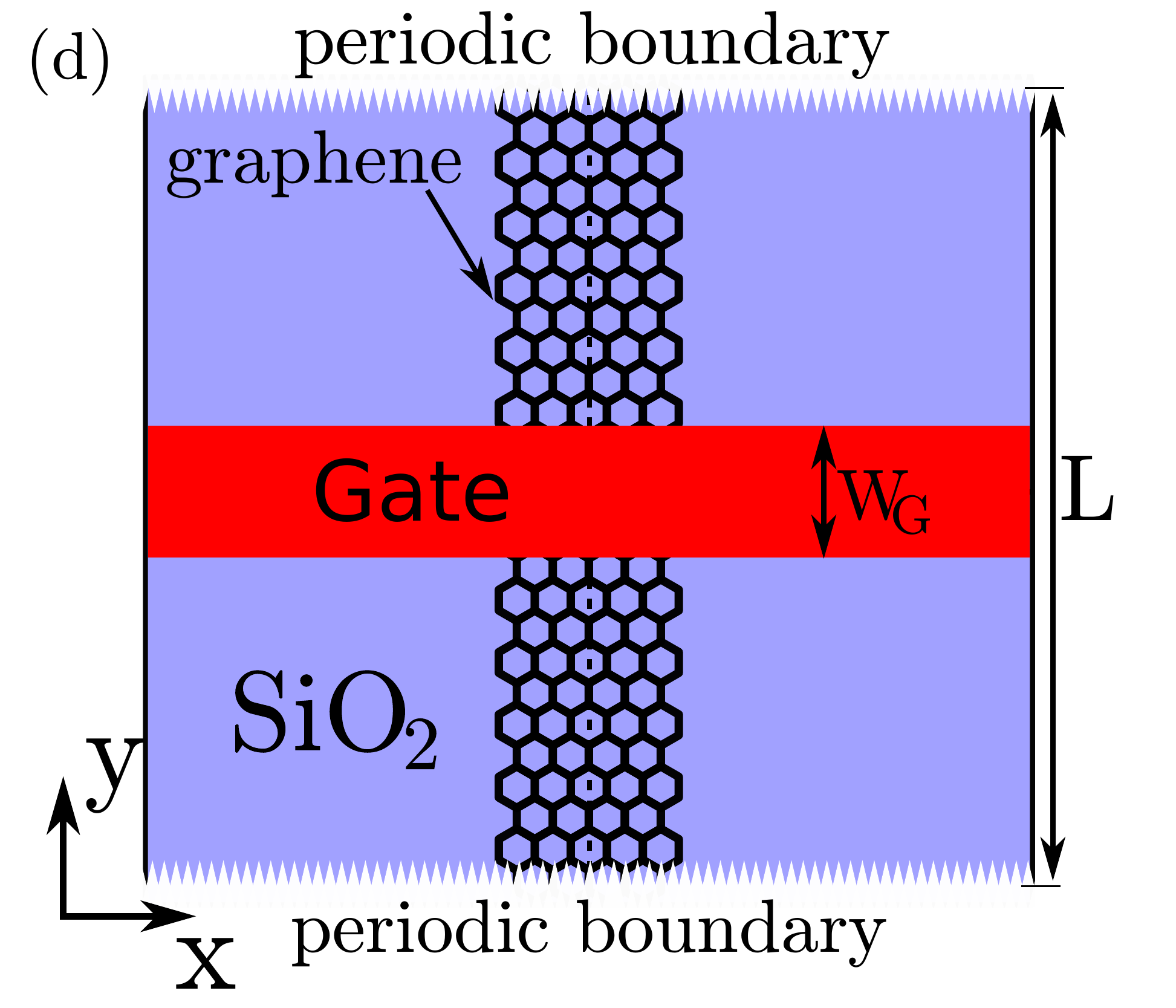}
 \caption{Schematic drawing of the considered model. Graphene deposited on the SiO$_2$ substrate with $ \epsilon=3.9$. The tip is located at the height of $z_{Tip}$ above the graphene.
 Metallic bottom gate is grounded. The computational box used for calculations is marked by the thin dotted lines, the height of the box above the structure equals $z_{Vac}=40$ nm. The width of the computational box is ten times bigger than the
 graphene for the normal component of the electric field to vanish at the edges. In (a) and (b) we show the side and top views of the system with point tip. In (c) we show the side view of the model with extended cone. The cone base is located at $z_C=10$ nm with radius $r_C=5$ nm. Finally, in (d) we show the top view of the model with a flat metallic gate of width $W_C=6$ nm.
 }

\label{Struct1}
 \end{figure}

We consider (Fig. \ref{Struct1}) a metallic and semiconducting armchair nanoribbon of width  $\approx 12 $ nm ($98$ carbon atoms across the channel for the metallic ribbon and 97 atoms for the semiconducting one). 
On the bottom of the system we place the metallic back gate (Fig. \ref{Struct1}(a,c)) which is covered by an insulating  SiO$_2$ layer $z_{Bg}=7 $ nm thick. 
The graphene is deposited on top of the insulator ($z=0$).
For the top gate we use three models of the tip: (i) a point-like tip that occupies a single cell (Fig. \ref{Struct1}(a,b)) of the finite difference mesh ($\Delta x=\Delta y=0.3$ nm and $\Delta z=0.2$ nm), (ii) an extended  cylindrical electrode ended with a cone tip \ref{Struct1}(c)) and (iii) a rectangular gate perpendicular to the ribbon.
The tip is placed above the graphene layer at the center of the ribbon, and the rectangular gate crosses the computation box symmetrically \ref{Struct1}(d)).

\begin{figure}[htbp]
\includegraphics[width=0.48 \textwidth]{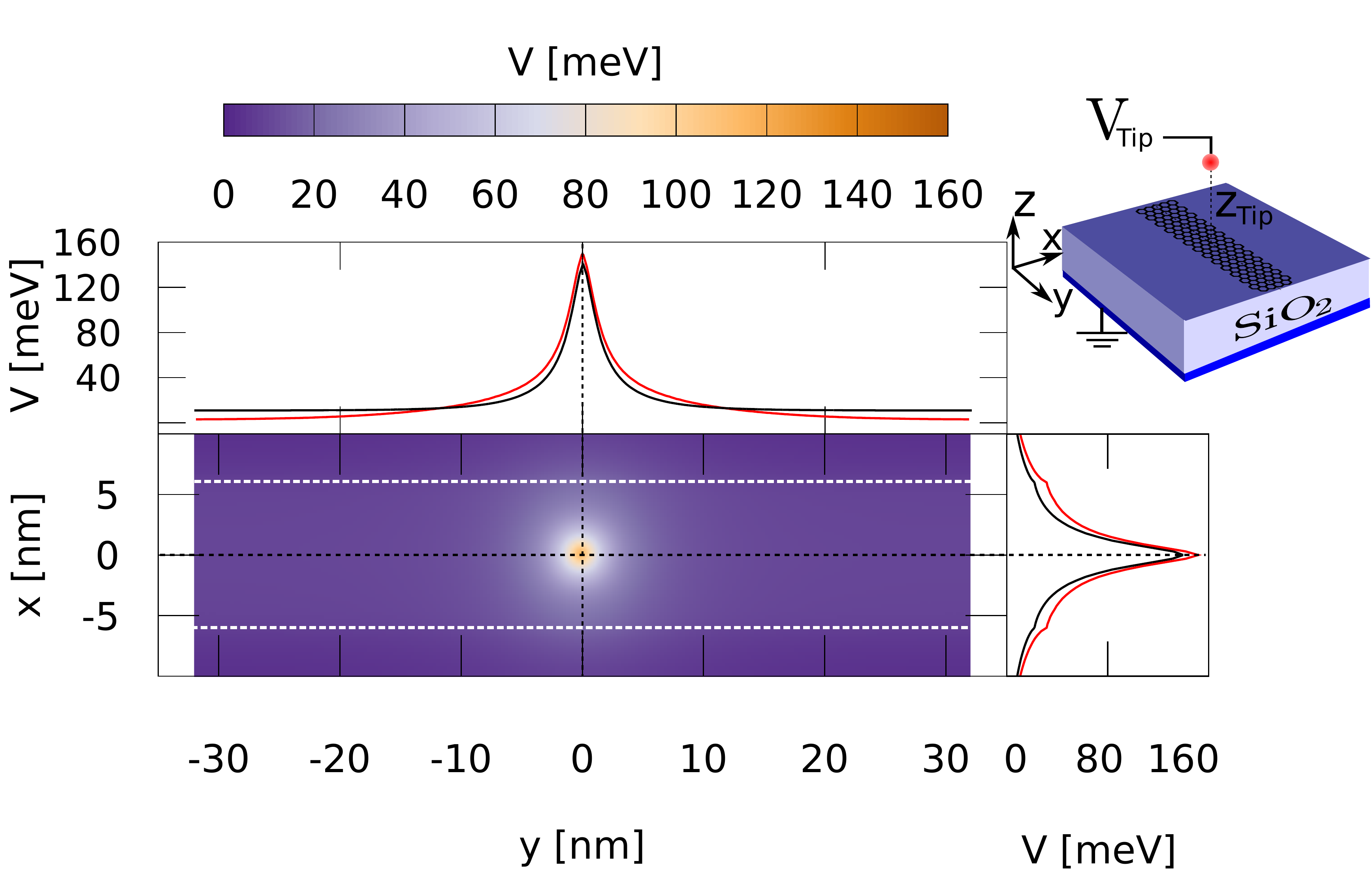} 
\caption{In the upper (right) plot we show potential energy along the $y$ axis for $x=0$ and $z=0$ ($x$ axis for $y=0$ and $z=0$) respectively. The black (red) line represents the metallic (semiconducting) nanoribbon respectively.
In the middle plot we show 2D distribution ($z=0$) for the metallic ribbon. The results are obtained for a point tip (see inset) located $1$ nm above the graphene flake with a potential $10$ V.
The dashed, white lines in the 2D distribution plot show the edges of the nanoribbon.
}
 \label{ppo}
\end{figure}

\begin{figure}[htbp]
\includegraphics[width=0.46 \textwidth]{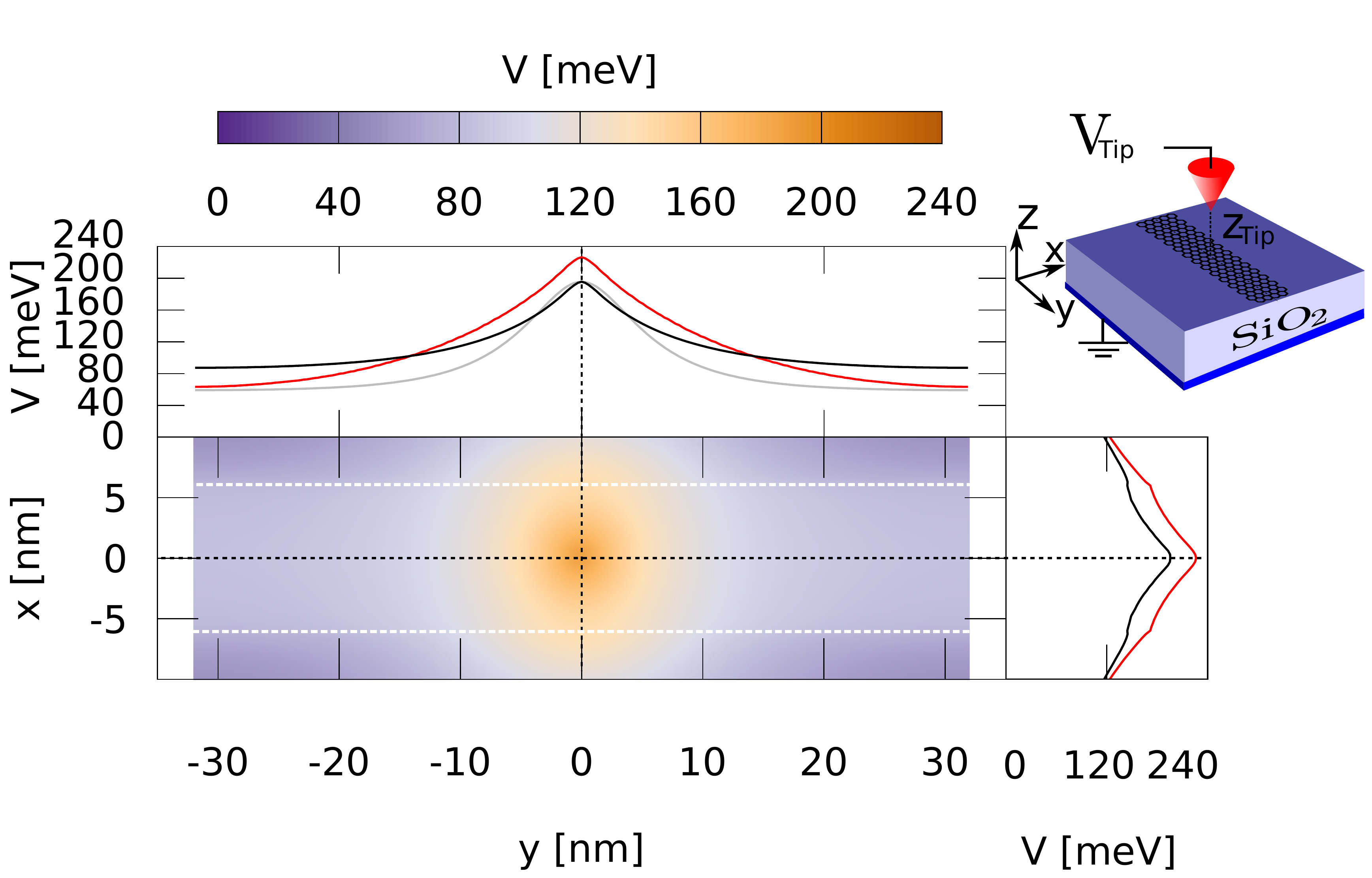} 
\caption{ (a) The same as Fig. \ref{ppo} but obtained for a metallic cone (see inset). 
The cone apex is located $1$ nm above the graphene flake ($z=1$ nm). The cone base is located at $z=10$ nm and its radius is equal to $5 $ nm. Above $z=10$ nm the cone becomes a cylinder with radius $5$ nm.
The cone and cylinder potential is set at $1.5$ V.
In the upper plot we also plot a gray line that represents the potential energy profile obtained for a point tip located $5$ nm above the graphene flake with a tip potential $88$ V.
}
 \label{ppo1}
\end{figure}

\begin{figure}[htbp]
a)\hspace{-0.5cm}\includegraphics[width=0.25 \textwidth]{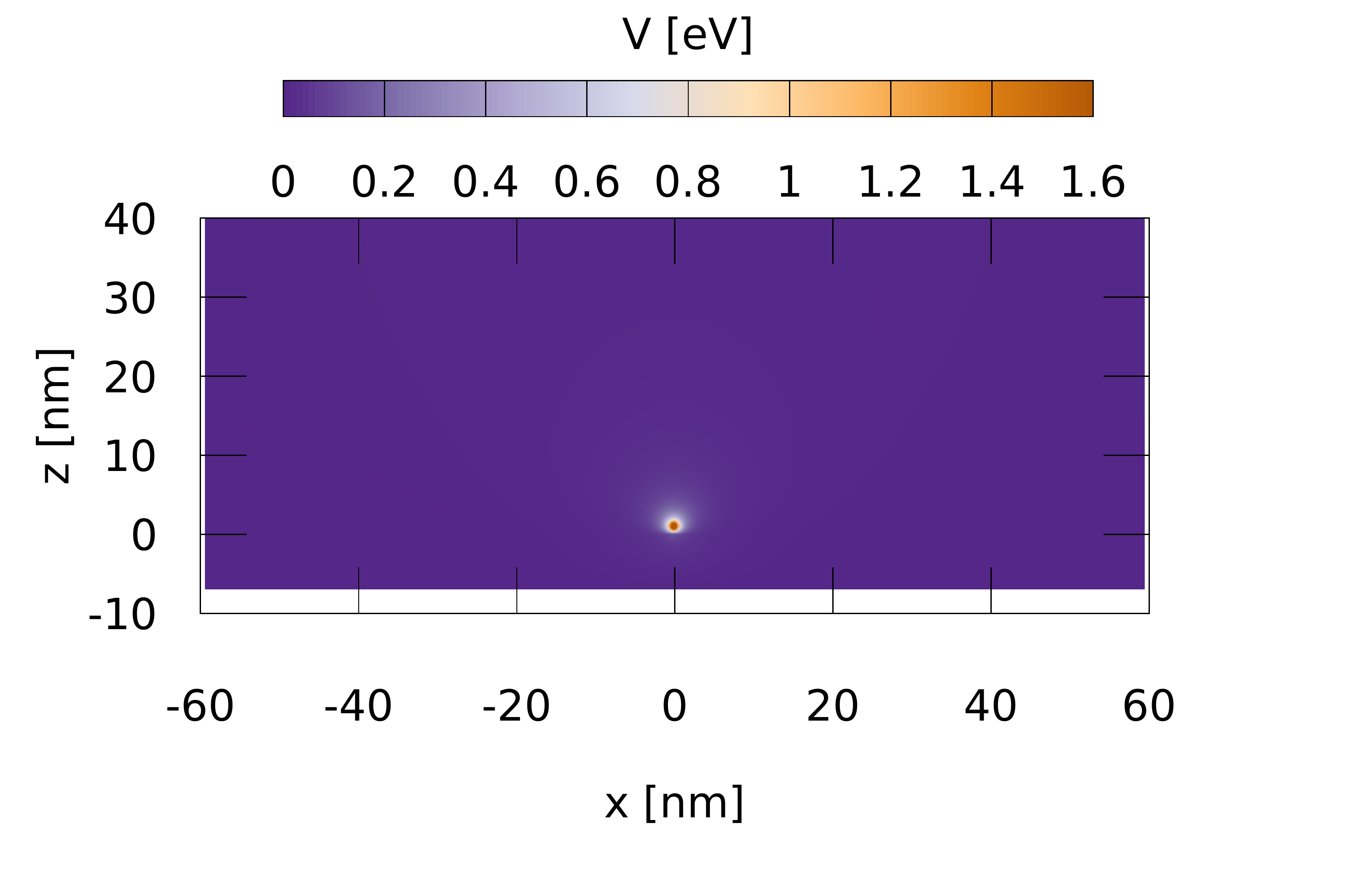} 
b)\hspace{-0.5cm}\includegraphics[width=0.25 \textwidth]{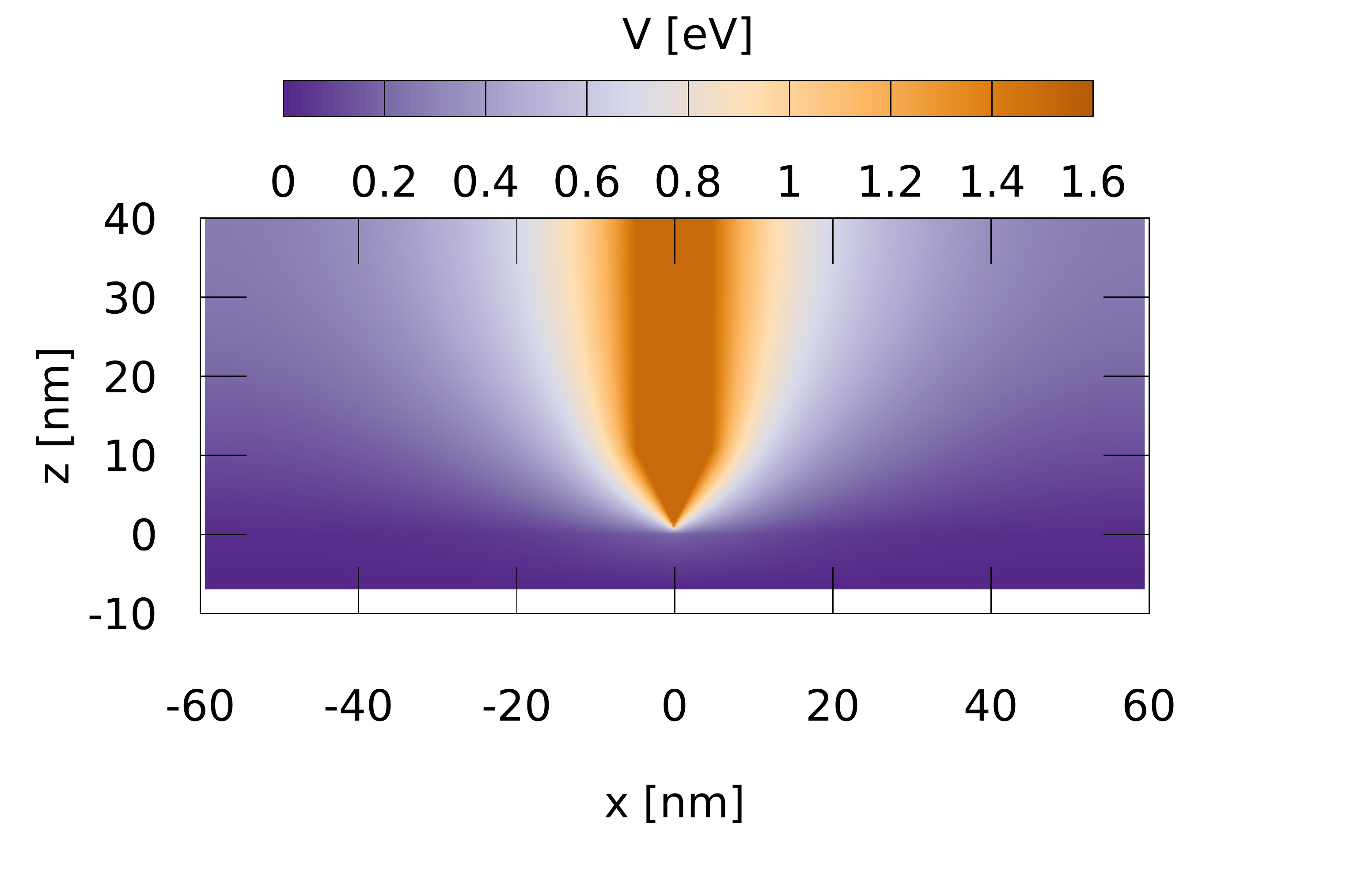} \hspace{-0.5cm}
\caption{The $x-z$ cross sections ($y=0$) of the total potential energy obtained for a point tip (a) and metallic cone (b). The color scale is the same on both figures.}
\label{xz}
\end{figure}

\begin{figure}[htbp]
\includegraphics[width=0.48 \textwidth]{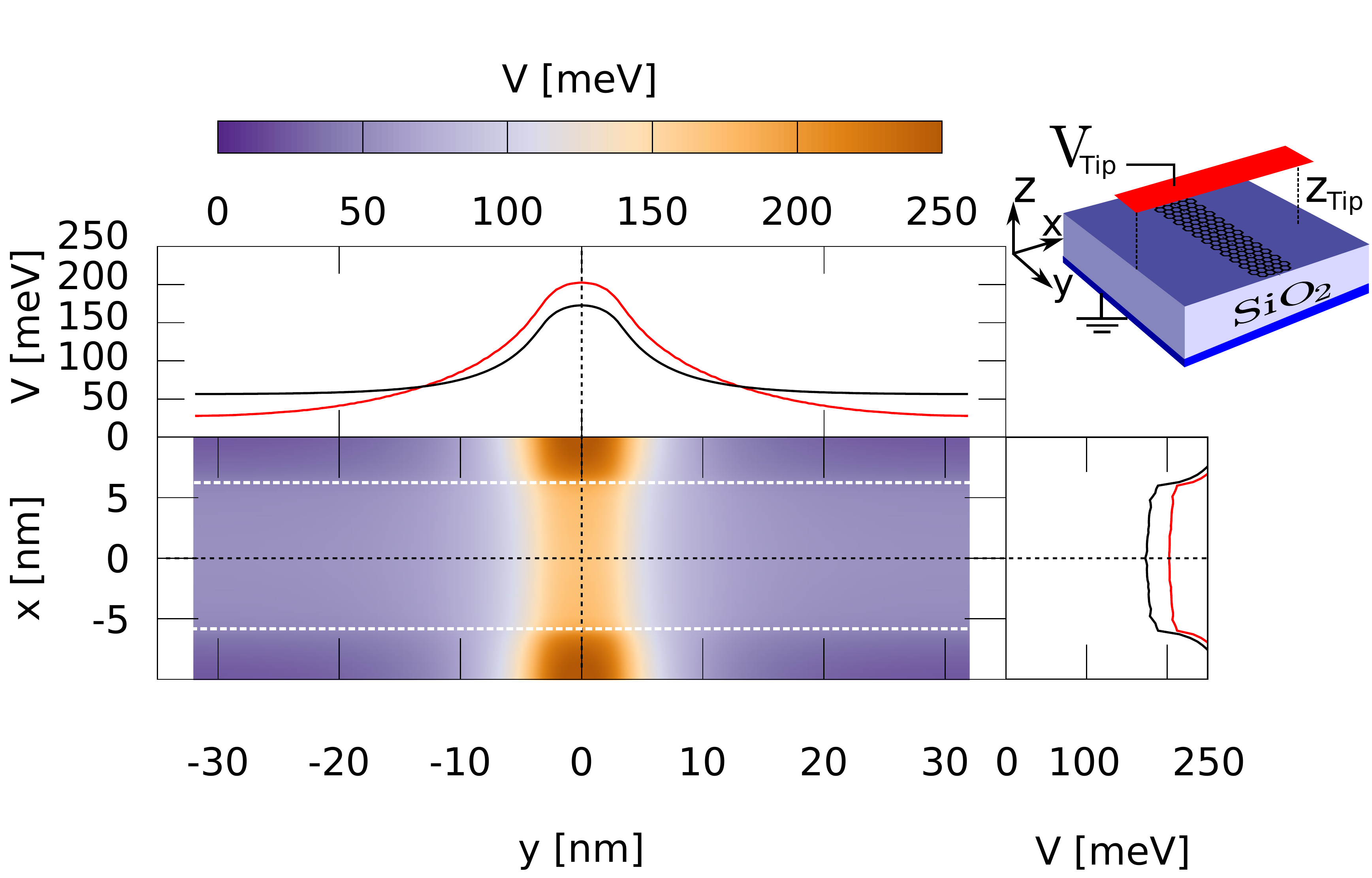} 
\caption{The same as Fig. \ref{ppo} but obtained for the flat top gate placed $z_{Tip}=1$ nm above the graphene flake. The gate potential is $V_{Tip}=0.5 $ V and the
gate width is equal $6$ nm.} 
\label{ppo2}
\end{figure}

For graphene we apply the standard tight-binding Hamiltonian form for the $\pi$ electrons on the $p_z$ orbitals
\begin{equation}
  \hat{H}_{DFT}=\sum_{i,\sigma}V_{i,\sigma} \hat{c}^{\dagger}_{i,\sigma} \hat{c}_{i,\sigma}+\sum_{ i,j,\sigma} t_{ij}\left( \hat{c}^{\dagger}_{i,\sigma}\hat{c}_{j,\sigma} +h.c \right ), \\
  \label{Hone}
\end{equation}
where $\hat{c}^{\dagger}_{i,\sigma}$ ($\hat{c}_{i,\sigma}$) is the creation (annihilation) operator of an electron at $i$-th atom with $\sigma$ spin and  $t_{ij} = -2.7$ eV \cite{CastroNeto2009} for all nearest ions and $0$ elsewhere.
The wave function for the $k$ -th eigenstate with $\sigma$ spin is defined as a linear combination of $p_z$ atomic orbitals
\begin{equation}
 \psi^k_{\sigma}(\mathbf{r})=\sum_{i}^{N}c^{k}_{i,\sigma} p_z(\mathbf{r-r}_i),
 \label{ffal}
\end{equation}
with orbitals $p_z(\mathbf{r-r}_i)$ centered at the carbon atom positions.
The interaction between electrons is taken into account by the Kohn-Sham potential of the form
\begin{equation}
\begin{split}
 V_{\sigma}(\mathbf{r})=&V_{p}(\mathbf{r})+V^{\sigma}_{xc}(n^{\uparrow},n^{\downarrow}), \\
 &V_{p}(\mathbf{r})=V_{ext}(\mathbf{r}) +V_{H}(\mathbf{r},n)+V_{I}(\mathbf{r},n_I),
 \label{potW}
\end{split}
 \end{equation}
  where $V_{ext}(\mathbf{r}),V_{H}(\mathbf{r},n)$ and $V_{I}(\mathbf{r})$ are external potential, Hartree potential and ion potential, respectively. 
  The exchange-correlation potential $V^{\sigma}_{xc}(n^{\uparrow},n^{\downarrow})$ is taken in the Perdew-Zunger form \cite{Perdew}.
  To proceed, we need to define electron-density $n(\mathbf{r})$ and ion-density, first one has form
  \begin{equation}
   \begin{split}
    &n(\mathbf{r})=n^{\uparrow}(\mathbf{r})+n^{\downarrow}(\mathbf{r}), \\
&n^{\sigma}(\mathbf{r})=\sum_{j=1}^{N}f(E_j,E_F,T)|\psi^{\sigma}_j(\mathbf{r})|^2,
  \label{gestoscf}
   \end{split}
  \end{equation}
where $f(E_j,E_F,T)$ is the Fermi-Dirac distribution function with orbital energy $E_j$ and Fermi energy $E_F$ in the $T$ temperature. 
The ion-density is defined as
\begin{equation}
 n_I(\mathbf{r})=\sum_{i}^{N} \delta(\mathbf{r-r}_i) .
 \label{gestoscI}
\end{equation}

We obtain the eigenstates of  Hamiltonian (\ref{Hone}) self-consistently using the Broyden \cite{Broyden} mixing scheme. 
The calculations were performed for a finite temperature of $T=5$ K, to achieve convergence.
For the Poisson equation, we use the finite difference method with the cubic cells.
For that reason the charge density defined within  the graphene ribbon is effectively
spread over a  finite thickness.
The external potential is defined by the metallic bottom gate and the metallic point tip above the structure (see. Fig. \ref{Struct1}). To calculate the entire potential of the system we solve the Poisson equation $-\nabla \left (\nabla\epsilon(z) V_p\left(\mathbf{r} \right) \right) =4\pi\rho\left(\mathbf{r} \right)$.
The applied finite difference approach is explained in the Appendix.

For the studies of the transport near the Dirac point we take the total potential that is derived for the charge neutral system. For the charge neutrality point the potential is independent of the spin. 
 For the potential on the bottom gate and on the tip we apply Dirichlet boundary conditions for fixed electrostatic potential. We set the bottom gate at $0$ (grounded) potential and vary the tip voltage. 
In order to simulate an infinite ribbon for evaluation of the potential profile in the $y$ direction we consider periodic boundary conditions with a box length equal to $L\approx 64 $ nm,
with about 30 thousand  carbon atoms inside the periodically repeated cell.
We apply the Dirichlet boundary conditions for the bottom face of the box, where the 
backgate is applied. For the other  faces we assume that the normal component of the field vanishes.
On the side faces perpendicular to $y$ axis this condition is justified by the periodic boundary conditions. 
On the side faces perpendicular to the $x$ axis the condition is justified by a large distance
from the tip, and -- for the rectangular top gate -- by the layer structure of the system.
At the top face of the computational box the vanishing of the electric field for 
the floating point-like tip and for the rectangular gate is justified by the charge neutrality
of the entire system. At a large distance from all the space charges the electric field, 
in particular its normal component vanishes. For the cone tip, the Neumann boundary condition
at the top side is justified by the radial character of the potential induced by the cylindrical gate.   
The conditions for the top face are justified at a large distance from the graphene layer 
so the computation box needs to be large enough.
To find an appropriate height of the computational box we calculate the potential energy profile 
on the graphene layer. The dependence is weak and  we set $z_{Vac}=40 $ nm as the value for which the convergence is reached.
For the lateral size of the box we find that it is sufficient to set the side length $W=120$ nm, i.e., ten times wider than the ribbon (Fig. \ref{Struct1}(b)).
The potential at the graphene ribbon at the end of the box is set as a reference energy level for the transport calculations.

 The size of the present model system is limited by the numerical complexity of the atomistic approach. Scaling approaches have been introduced 
for the tight binding models with the increase of the carbon-carbon distance that is compensated by the modification of the hopping parameters \cite{Rickhausprl}.
In the present paper we keep the carbon-carbon distance unchanged and the considered system should be treated as a model with physical dimensions reduced by a factor of $\simeq 10$
of the actual device. Accordingly, for
 the transport modeling we use the potential that is obtained for the tip at 1 nm above the graphene plane.

Fig. \ref{ppo} shows the results for the floating point-like model of the tip for 
the metallic (black) and semiconducting (red) ribbon. The potential is screened
more effectively by the electron gas in the metallic ribbon. 
We can see that the screening by the ribbon produces an anisotropy in the effective potential,
with the maximum that is elongated perpendicular to the ribbon axis. 
This anisotropy and the dependence of the effectiveness of the screening 
on the type of the ribbon are also found for the cone-model of the tip plotted in Fig. \ref{ppo1}. 
The cone model of the extended tip produces wider maximum than the point-like model. 
For illustration a cross-section of the potential for the tip models is given in Fig. \ref{xz}.
For the extended top-gate electrode [Fig. \ref{ppo2}] the stronger screening by the metallic
ribbon is evident.

\subsection{Transport}
 
The effective potential evaluated by the tip, including the screening by all the occupied single-electron energy levels below the Fermi level,  
is calculated in the absence of the external magnetic field. 
For the transport at the Fermi level we account for the external magnetic field modifying the hopping parameters of Eq.~(\ref{Hone}) with
 the Peierls phase $t\rightarrow t_{ij}=t \exp\left( \frac{2\pi i}{\phi_0} \int_{\mathbf{r}_i}^{\mathbf{r}_j} \mathbf{A}\cdot d\mathbf{l} \right)  $, where $\mathbf{A}$ is the vector potential. We use the gauge appropriate for the terminals using the approach proposed in Ref.~\onlinecite{Baranger1989}. We take $\mathbf{A_0}=(By,0)$ and apply a gauge transformation $\mathbf{A}=\mathbf{A_0}+\mathbf{\nabla}[f(x,y)m(x,y)]$ with $f(x,y)=-xyB$, and $m(x,y)$ a smooth step-like function that is 0 in the ribbon interior  interior where the probes attached, and 1 for $|y|>$24 nm .
 We introduce
the spin Zeeman effect,
\begin{eqnarray}
  H'&=& H + \frac{1}{2}g\mu_B B \sum\limits_{i,\sigma } (\hat\sigma_z)_{\sigma,\sigma} c_{i\sigma}^\dagger c_{i\sigma}   , 
\label{eq:dh}
\end{eqnarray}
where $\mu_B$ is Bohr magneton, $\hat\sigma_z$ is the $z$ Pauli matrix, and $g=2$.

\begin{figure}[tb!]
 \includegraphics[width=\columnwidth]{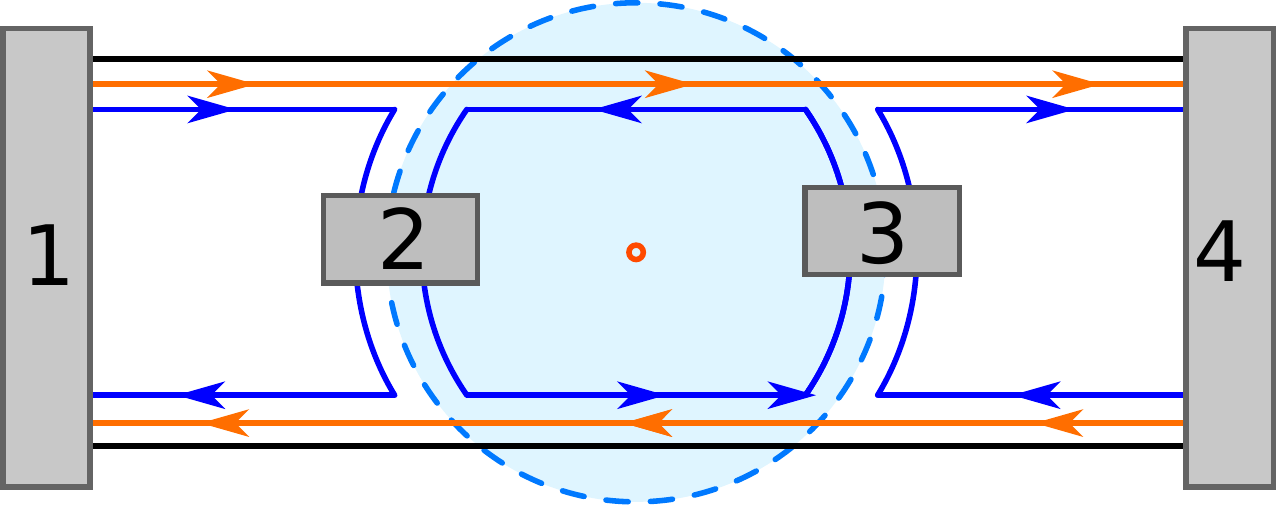}
  \caption{The scheme of the considered system. The blue area indicates the region with inverted charge of the carriers 
with the $n-p$ junction marked by the dashed line. The net current flows from lead 1 to lead 4. The leads 2 and 3 
are the B\"uttiker probes implemented as voltage terminals.} \label{system}
\end{figure}

For evaluation of the transmission probability, we use the wave function matching (WFM) technique \cite{Kolacha}. The transport calculations are performed for zero temperature. The leads feeding and draining current are considered semi-infinite. The partial transmission probability is evaluated as
\begin{equation}
T^{kl}_{m,\sigma} = \sum_{ n,\sigma' } |t^{kl}_{mn,\sigma\sigma'}|^2,
\label{eq:transprob}
\end{equation}
where $t^{kl}_{mn,\sigma\sigma'}$ is the probability amplitude for the transmission from the mode $n$ with spin $\sigma'$ in the input lead $l$ to mode $m$ with spin $\sigma$ in the output lead $k$.
The Hamiltonian (\ref{eq:dh}) is spin-diagonal. 
We evaluate the conductance from lead $l$ to $k$ as $G_{kl}={G_0}\sum_{m,\sigma} T^{kl}_{m,\sigma}$, with $G_0={e^2}/{h}$. For a single spin $G_{kl,\sigma}={G_0}\sum_{m} T^{kl}_{m,\sigma}$.
In our model the dephasing is defined by the coupling of the sample to the  virtual probes $t_B$. 
For fully coherent transport with the virtual probes detached  ($t_B=0$)  the two-terminal conductance is simply given by $ G_{j\rightarrow i} =G_{ij} $.

To take into account the dephasing, we use the B\"uttiker virtual probes \cite{Buttiker1988} technique. 
The idea \cite{Buttiker1988} is to introduce leads with zero net current fed to the system, i.e. an electron that enters the probe, gets to the reservoir and comes back to the system with a randomized phase.  Technically, one solves the transport problem at the Fermi level for a model multiterminal device. 
In general, for a system with $N$ terminals, 
the conductance matrix $\boldsymbol{\mathcal{G}}$ is constructed, which translates the voltage to currents
$ \boldsymbol{I}=\boldsymbol{\mathcal{G}}\boldsymbol{V} $, with $I_i$ being the current flowing in the $i$th terminal and $V_i$ the voltage measured at the $i$th terminal. The matrix elements are
calculated as 
\begin{equation}
\mathcal{G}_{ii} = \sum\limits_{j=1,j\ne i}^{N} 
G_{ij}
\label{eq:Gdiag}
\end{equation} 
with
\begin{equation}
\mathcal{G}_{ij} = - G_{ij}.
\label{eq:Goffdiag}
\end{equation} 
The currents in the leads are related by the Kirchhoff's law, thus the $\boldsymbol{\mathcal{G}}$ matrix as given by Eqs. (\ref{eq:Gdiag}) and  (\ref{eq:Goffdiag})  is singular.
After  elimination of the current in the $N$th terminal $I_{N}$ one obtains an invertible $(N-1)\times (N-1)$ $\boldsymbol{\mathcal{G}}$ matrix. The resistance matrix is then evaluated by $\boldsymbol{\mathcal{R}}=\boldsymbol{\mathcal{G}}^{-1}$. 

For the system with equilibration, we use two B\"uttiker probes, so the system has $N=4$ terminals in total as in Fig. \ref{system}. 
The net current flows from terminal 1 to terminal 4. Leads 2 and 3 are B\"uttiker voltage terminals with zero net current.  The two-terminal conductance is given by
\begin{equation}
G_{i\rightarrow j} = \frac{I_i}{ V_{i}-V_j }.
\label{eq:Gij}
\end{equation} 
 For the labeling of the terminals shown in Fig.~\ref{system}, we set the potential of the lead 4 as a reference $V_4=0$. The two-terminal conductance is then given by
\begin{equation}
G_{1\rightarrow 4} = \frac{I_1}{ V_{1}-V_4 } =\frac{I_1}{ V_{1} } = \frac{1}{ \mathcal{R}_{11} }.
\label{eq:G14}
\end{equation} 
For each of the terminals we solve the scattering problem with the WFM method to obtain $G_{ij}$, fill in the $\boldsymbol{\mathcal{G}}$ matrix, and compute the conductance (\ref{eq:G14}). From now on we use the notation $G=G_{1\rightarrow 4}$ for the two-terminal conductance.

\begin{figure}[tb!]
 \includegraphics[width=0.48\columnwidth]{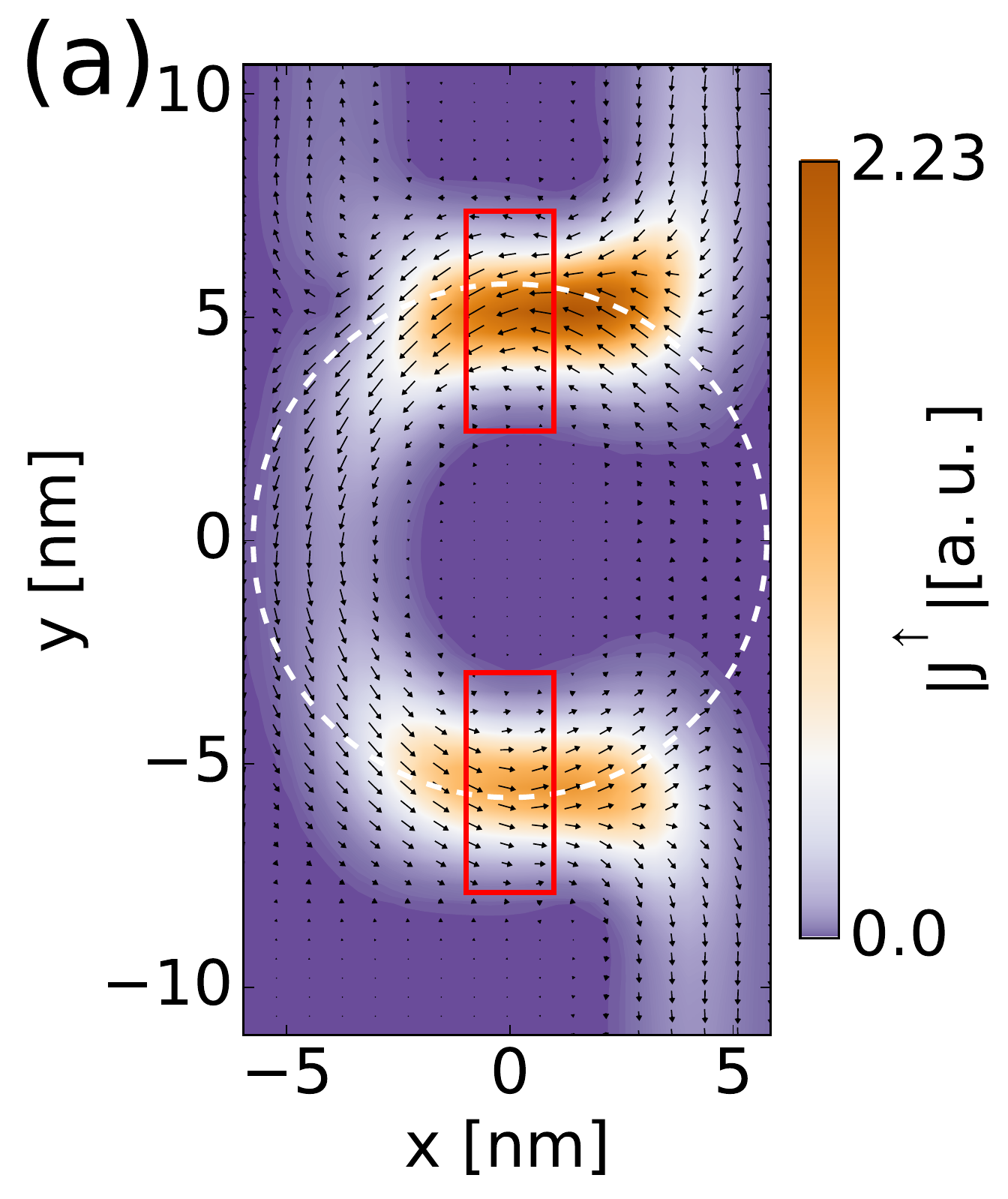}
 \includegraphics[width=0.48\columnwidth]{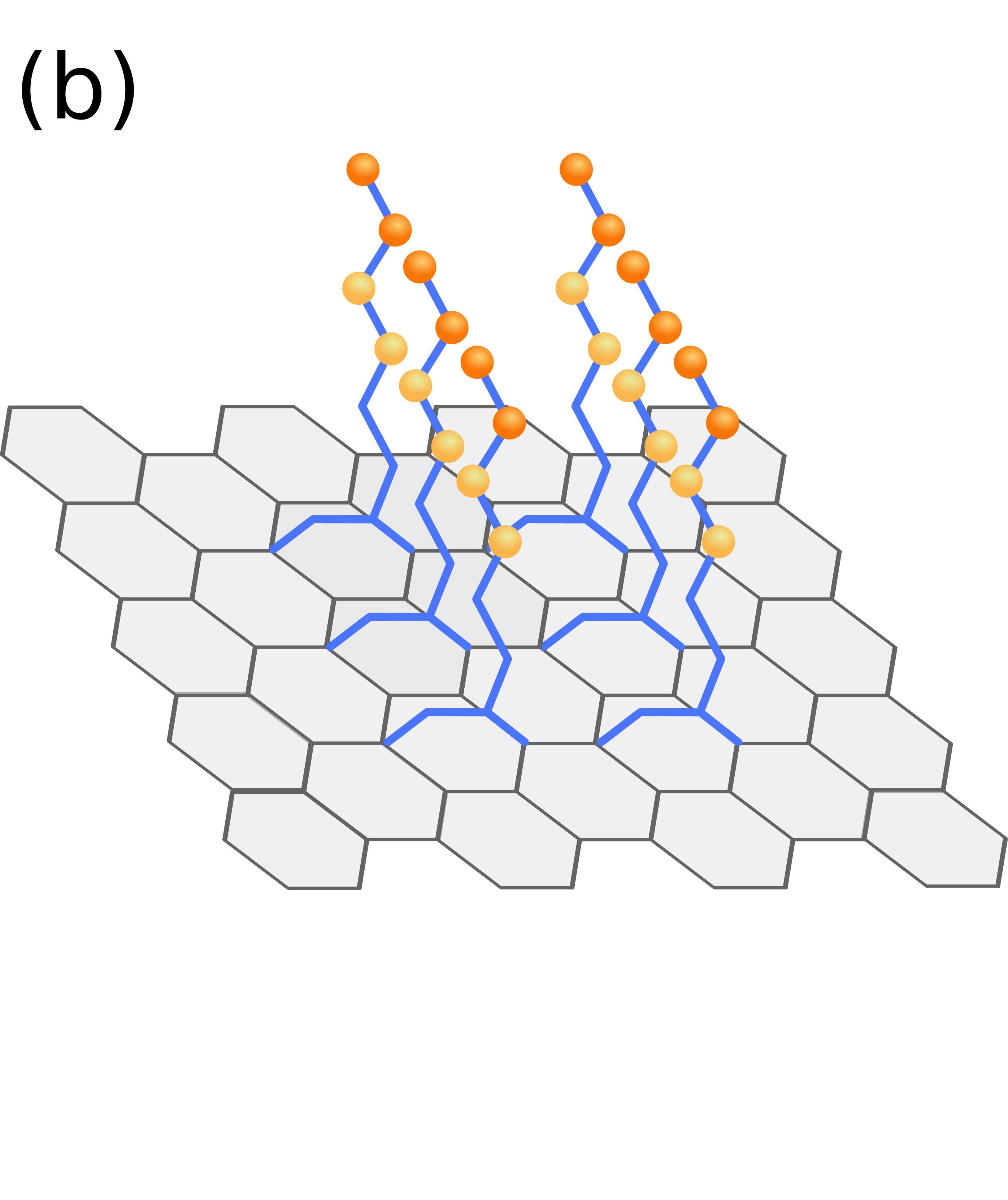}
  \caption{ (a) Current density for spin up electrons for $B=0.0038 \phi_0$ and $E_F=0.005$ meV. The electron is incident from the top. The white dashed circle is the $n-p$ junction. The red rectangles show the position and extent of the virtual probes. (b) The scheme of the B\"uttiker probes used in the calculations. The balls marked in yellow belong to a first elementary cell of the probe attached to the graphene and the orange balls for the second cell.
  } \label{systemBut}
\end{figure}

Let us describe the implementation of the B\"uttiker probes. 
For low Fermi energy the tip potential introduces an n-p junction to the ribbon.
The p region within the n medium  is either circular or it occupies the entire width of the ribbon. 
There are at most two bipolar junctions where the dephasing takes place. We connect voltage probes to the graphene plane near the ribbon axis at the position of the junction -- Fig.~\ref{systemBut}.
For the equilibration along the bipolar junctions in order to maximize the dephasing, the probe has to span across the junction (see red rectagles in Fig.~\ref{systemBut}(a)).
The applied probes consist of a set of one-dimensional zigzag chains connected to the graphene  as depicted in Fig.~\ref{systemBut}(b). 
The hopping between the graphene and the probe is $t_B$, which is a parameter that allows us to switch on or off the dephasing probes from the system.
Our probes  consist of 99 zigzag chains connected to a common reservoir. The position is dynamically set to the position of the bipolar junction, i.e. the center of the probe is where $E_F\pm \frac{1}{2}g\mu_B B =V(\mathbf{r})$ for spin up (down) electrons.
Note that the probes work effectively only when the system enters the quantum Hall regime and the current flows close to the edge and along the junctions and not in the entire width of the ribbon. 



\section{Results}

\subsection{Fully coherent transport}

\begin{figure*}[tb!]
 \includegraphics[width=0.32\textwidth]{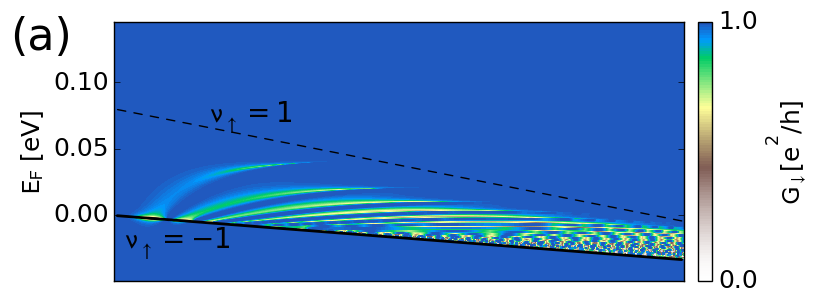}
 \includegraphics[width=0.32\textwidth]{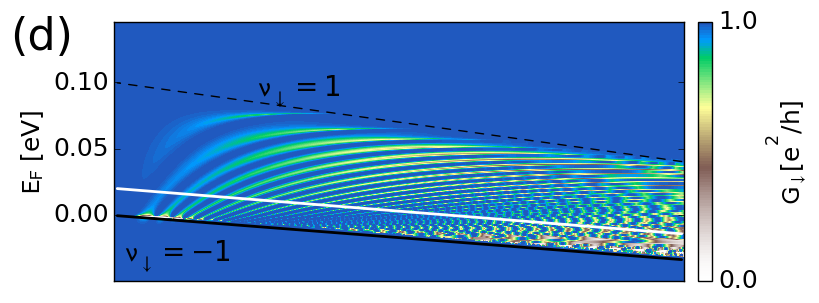}
 \includegraphics[width=0.32\textwidth]{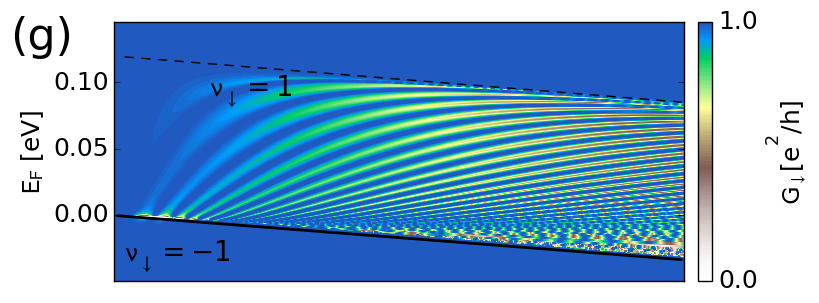}
 
 \includegraphics[width=0.32\textwidth]{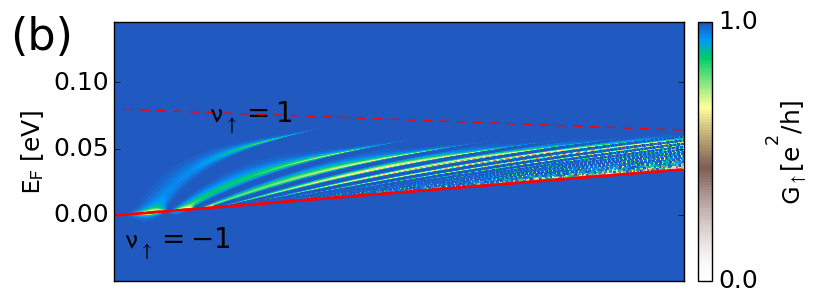}
 \includegraphics[width=0.32\textwidth]{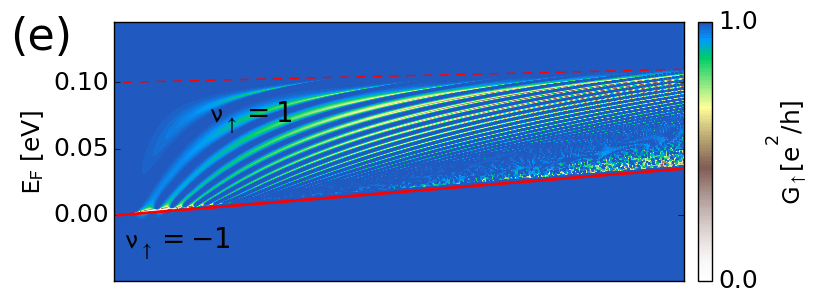}
 \includegraphics[width=0.32\textwidth]{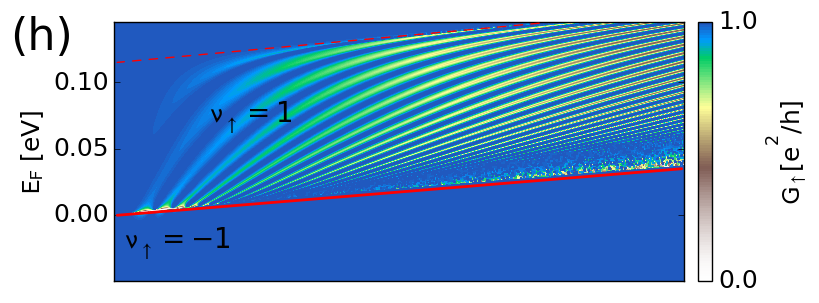}
 
 \includegraphics[width=0.32\textwidth]{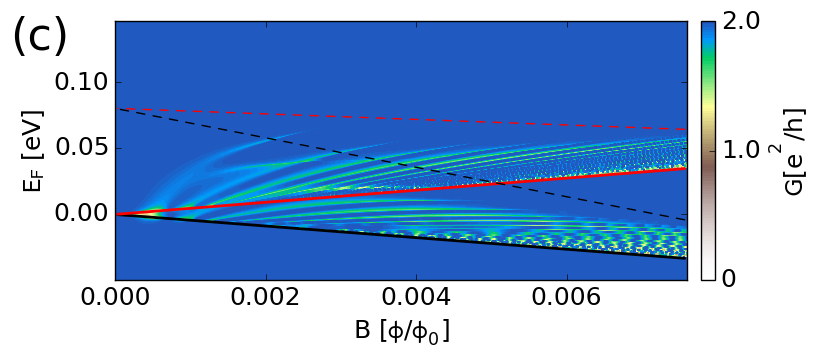} 
 \includegraphics[width=0.32\textwidth]{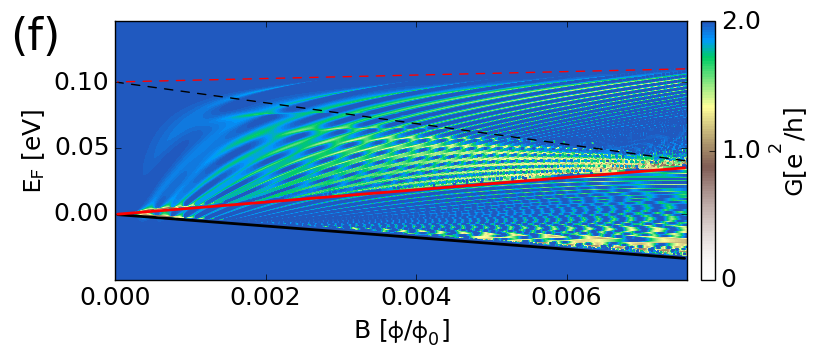} 
 \includegraphics[width=0.32\textwidth]{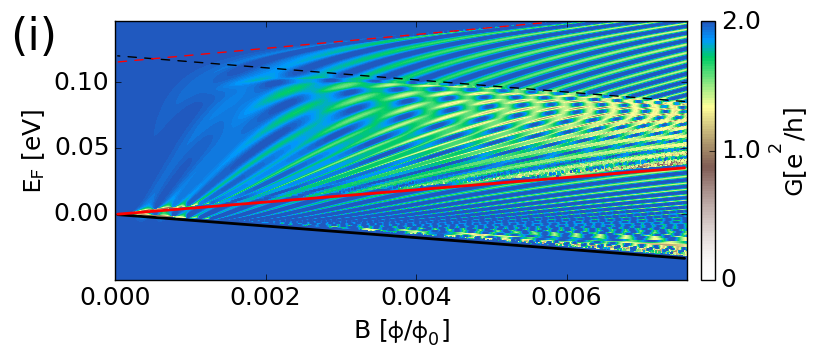}
  \caption{The fully coherent conductance in the metallic nanoribbon for (a,d,g) spin down, (b,e,h) spin up, (c,f,i) sum over both spin directions. The first column (a-c) corresponds to the point-like tip model,
(d-f) for the cone tip and (g-i) were obtained for a rectangular top-gate. 
The solid lines show the Dirac point, and the dashed lines 
 show approximately where the junction states decouple from the edge states. This decoupling for (a-f) occurs at the energies
where the junction is still present within the sample, i.e. for $E_F$ lower than the potential maximum. For (g-i) this line shows the potential maximum as for the flat gate the edge states cannot decouple from the junctions states.
 In (a,d,g) and (b,e,h)
the filling factor for a single spin is shown. The voltage for the point-like tip was $V_{Tip}=10$ V, for the cone-like tip $V_{Tip}=1.5$ V and for the flat gate $V_{Tip}=$0.5 V.
  } \label{koherentKr}
\end{figure*}

Let us first consider the fully coherent magnetotransport in the nanoribbons with tip-induced circular $n-p$ junction. Fig.~\ref{koherentKr} shows the spin down (a,d,g), spin up (b,e,h), and overall (c,f,i) conductance of a metallic nanoribbon, as a function of magnetic field through one graphene hexagon ($\phi=3\sqrt{3}a_{CC}^2 B$, with $a_{CC}=0.142$ nm, $\phi_0=\tfrac{h}{e}$) and Fermi energy. $\phi=\phi_0$ for $B=39.5\cdot 10^3$ T.
The conductance is generally close to 2$\tfrac{e^2}{h}$ with the exception
of conductance oscillations involving the circular $n-p$ junction. The Dirac point for both directions of spin shifts in opposite energy, as indicated by red and black thick solid lines in Fig.~\ref{koherentKr}. 

Figures \ref{koherentKr}(a-c) show the results for a potential profile calculated for a point-like tip. 
The $p$ island within the $n$ area   in Fig.~\ref{koherentKr} is formed above the solid line -- indicating the Dirac point. Above the threshold shown by dashed red (black) line for spin up (down) the oscillations vanish as the radius of the $n-p$ junction gets so small that the edge states no longer couple to the states around the junction. 

Figures \ref{koherentKr}(d-f) present the conductance for tip modeled as a cone. For such a potential profile the n-p junction is more extended, and therefore it can reach the edges of the ribbon for higher carrier energies than in the case of point-like tip. Thus the oscillations occur in a broader range of energies than for the results in Fig. \ref{koherentKr}(a-c). Still, the higher the magnetic field, the narrower the extent of the wave functions flowing in the ribbon, thus the coupling to the junction gets smaller.

In the Figures \ref{koherentKr}(g-i) the results for a flat electrode are shown. In this case the dashed line shows where the sum of Fermi and Zeeman energy exceed the maximal potential energy of the induced n-p junction.


The Aharonov-Bohm periodicity $\Delta B$ of conductance is related to the area $A$ of the effective $p$ island by the formula $ \Delta B = \tfrac{h}{eA} $. The area and the period are similar for the structures considered in Fig. \ref{koherentKr}.

\begin{figure*}[tb!]
 \includegraphics[width=0.49\textwidth]{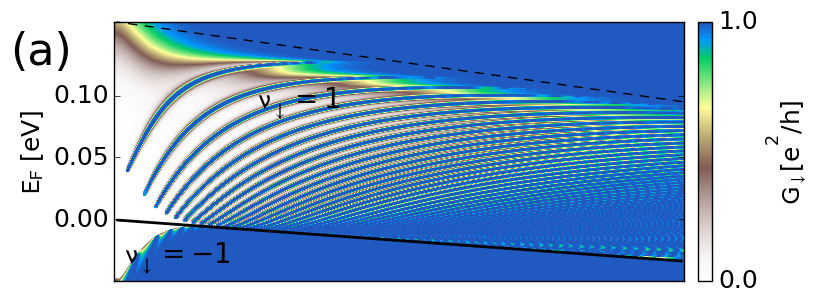}
 \includegraphics[width=0.49\textwidth]{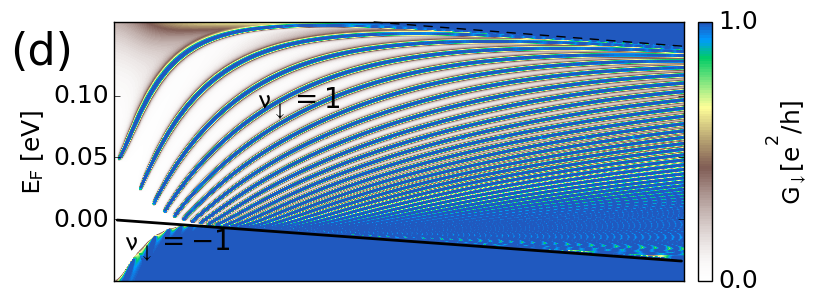}
 
 \includegraphics[width=0.49\textwidth]{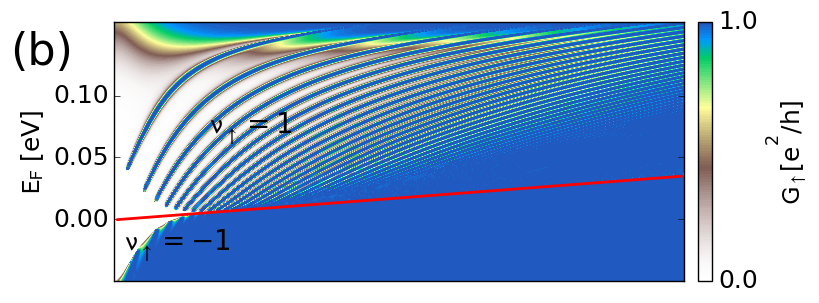}
 \includegraphics[width=0.49\textwidth]{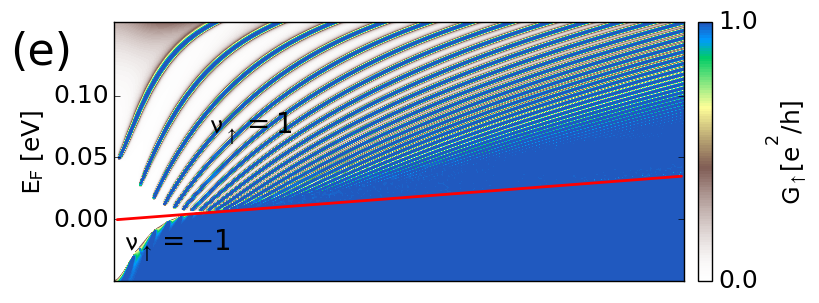}
 
 \includegraphics[width=0.49\textwidth]{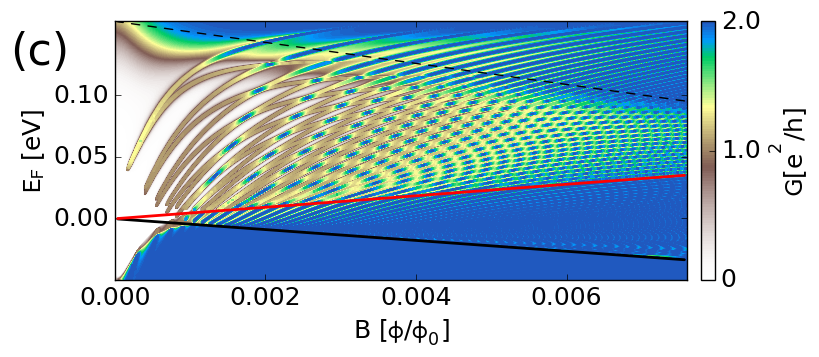} 
 \includegraphics[width=0.49\textwidth]{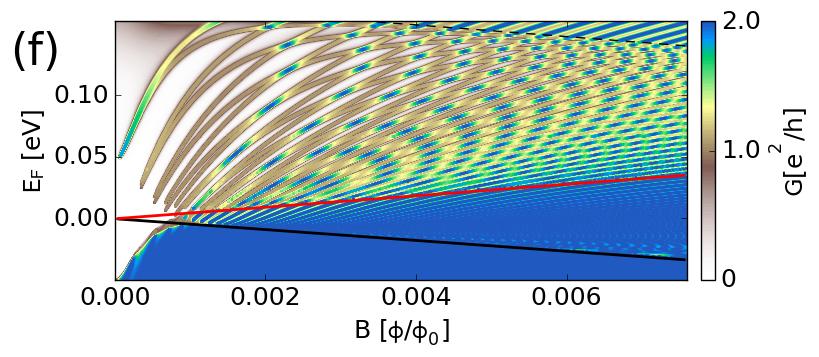}
  \caption{ Fully coherent conductance in the semiconducting nanoribbon for (a,d) spin down, (b,e) spin up, (c,f) sum over both spin directions. The solid lines in (a,b) and (c) show the Dirac point. Left column (a-c) corresponds to the cone tip and (d-f) to a rectangular top gate.
  } \label{koherentKrSem}
\end{figure*}

Figure \ref{koherentKrSem} presents the spin down (a,d), spin up (b,e), and overall (c,f) conductance of a semiconducting nanoribbon, for the potential due to a cone-shaped tip and for the flat gate potential. The difference with respect to the metallic nanoribbon is that the amplitude of the oscillations is much higher,
as previously reported in Ref. \cite{Mrenca2016}.

\subsection{Equilibration }

The conductance of a bipolar $n-p-n$ junction with equilibrated currents and  filling factors $\nu/\nu'/\nu$ in a spin-degenerate system is given by \cite{Ozyilmaz2007}:
 \begin{eqnarray}
 G= \frac{e^2}{h} \frac{|\nu'||\nu|}{2|\nu'|+|\nu|}.
\label{eq:Geq}
\end{eqnarray}
Here the filling factor is summed over spins.
For instance in a spin degenerate system with filling factors in the junction $+2/-2/+2$, the conductance with complete equilibration is $\tfrac{2}{3}\tfrac{e^2}{h}$. The channel mixing is only between channels of the same spin \citep{Amet2014,Weie2017}. For each of the spins the conductance is $G_\sigma= \tfrac{e^2}{h} \tfrac{|\nu_\sigma'||\nu_\sigma|}{2|\nu_\sigma'|+|\nu_\sigma|}$, giving $G_{\uparrow}=G_{\downarrow}=\tfrac{1}{3}\tfrac{e^2}{h}$, and the overall conductance is the sum $G=G_{\uparrow}+G_{\downarrow}$.

\begin{figure}[tb!]
 \includegraphics[width=1.\columnwidth]{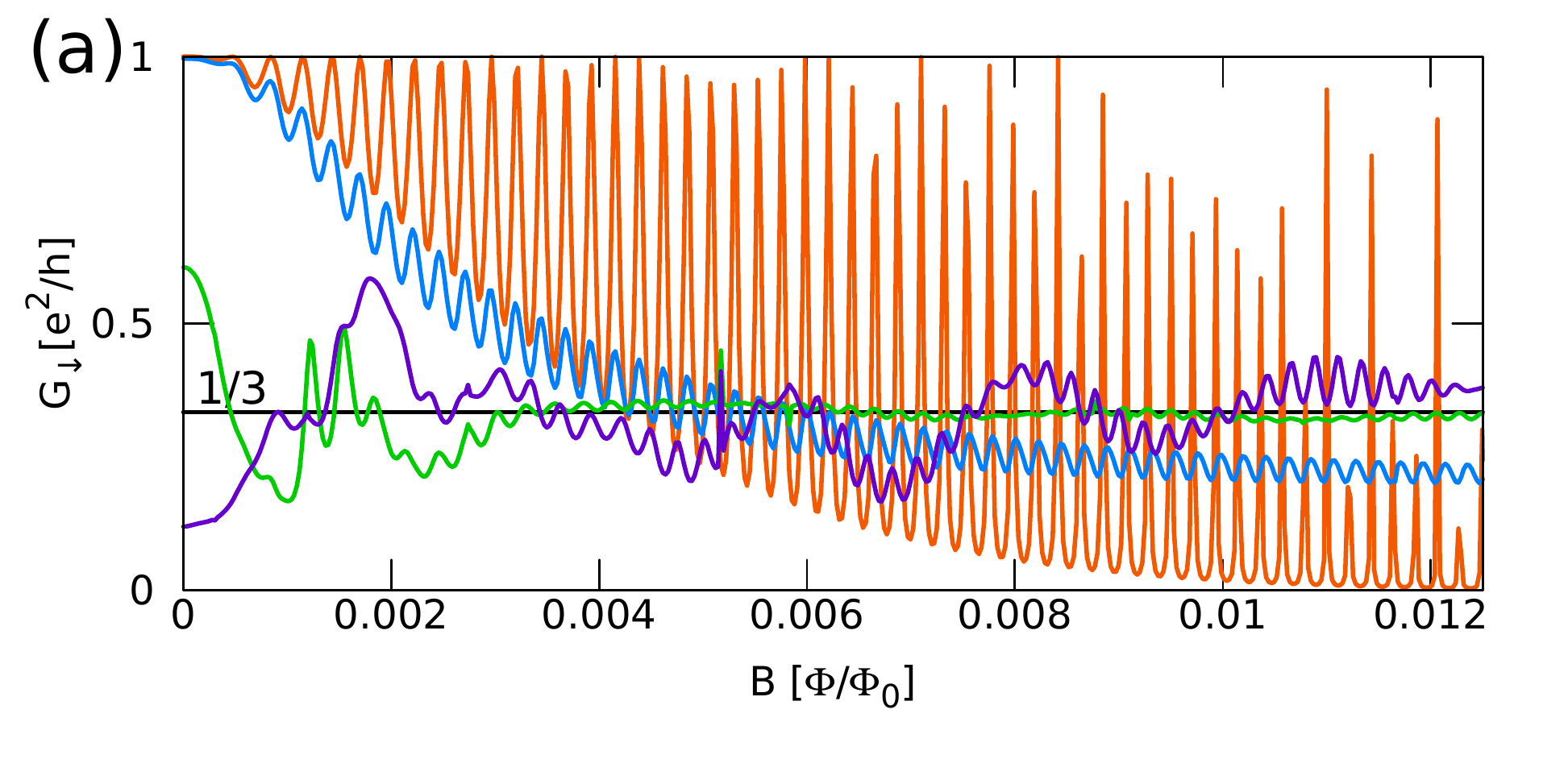}
 \includegraphics[width=1.\columnwidth]{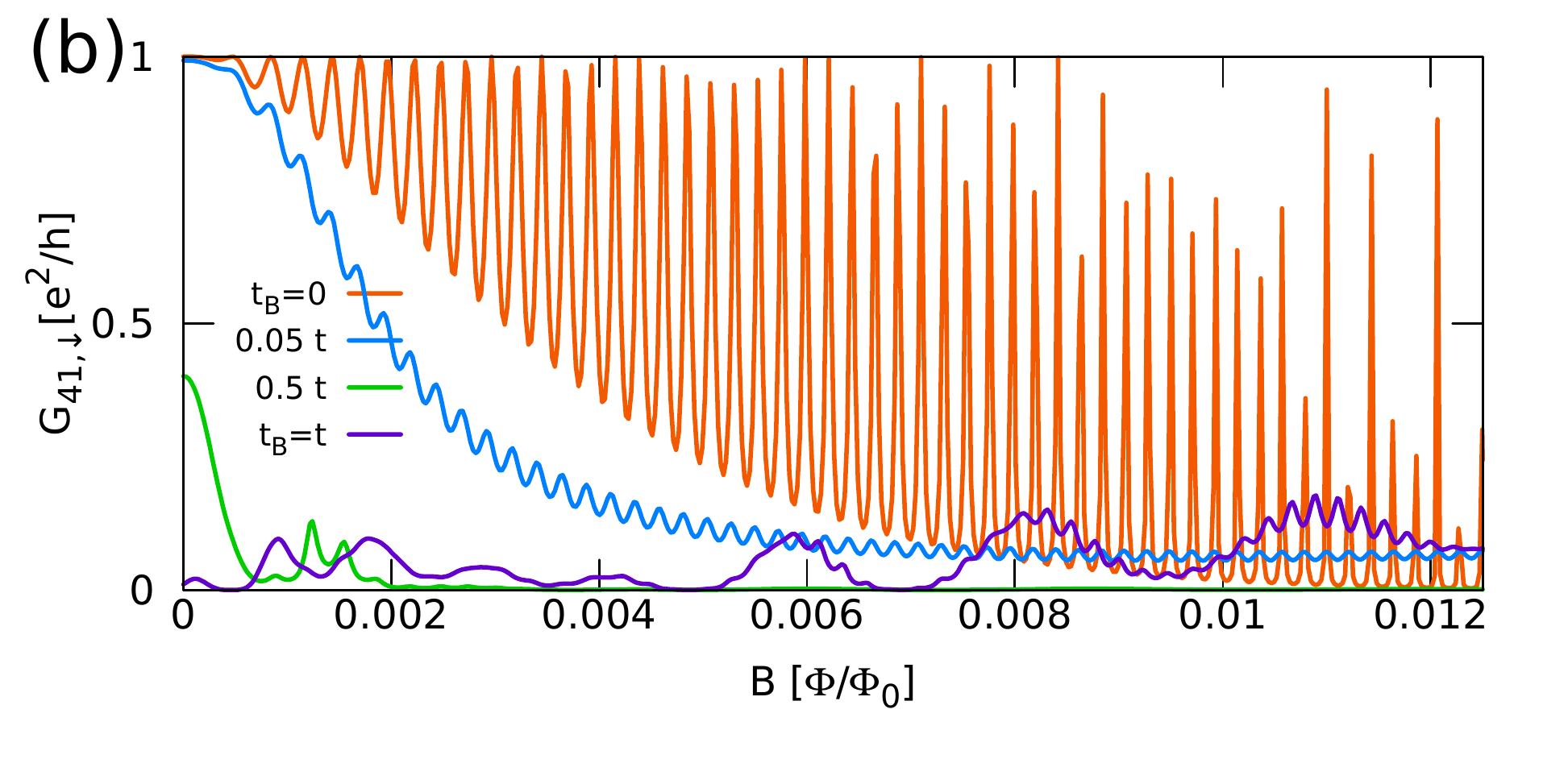}
  \caption{The cross section of conductance with dephasing by B\"uttiker probes in the wider nanoribbon along the white inclined line in Fig.~\ref{koherentKr} for (a) the two-terminal conductance and (b) the $G_{41}$ transmission for spin down Fermi level. 
  } \label{inkoherentKrCross}
\end{figure}

 Fig.~\ref{inkoherentKrCross}(a) shows the cross-section of Fig. \ref{koherentKr}(d) along the white inclined line for varied parameter $t_B$.
 Clearly the oscillations for $t_B=0$ have the highest amplitude, and for growing $t_B$ their amplitude is more and more suppressed. For $t_B=0.5 t$, at  high magnetic field limit,  the conductance saturates 
at $G_\downarrow\approx \tfrac{1}{3} \tfrac{e^2}{h}$. In Fig. ~\ref{inkoherentKrCross}(a) the conductance matrix element $G_{41}$  derived directly from the quantum scattering problem is presented for comparison. For $t_B>0$ the visibility of conductance oscillation is distinctly larger in the two-terminal conductance than in the matrix element, and the two-terminal conductance is non-zero for $t_B>0$ at high magnetic field, although the matrix element vanishes. The vanishing of $G_{41}$ coincides with the saturation of the two-terminal conductance at $\tfrac{1}{3} \tfrac{e^2}{h}$, since all the electrons flow into the virtual leads and do not reach the output lead. 

\begin{figure*}[tb!]
 \includegraphics[width=0.49\textwidth]{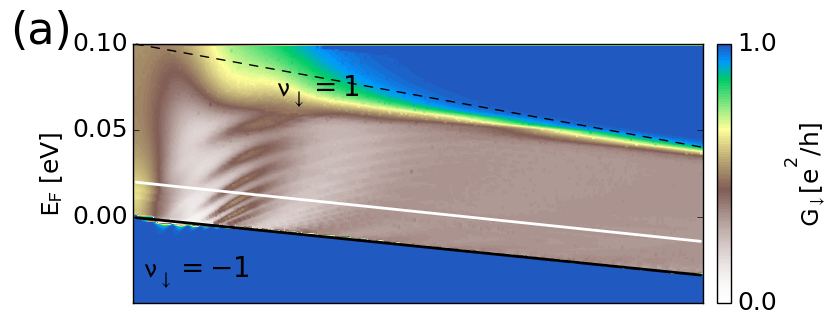}
 \includegraphics[width=0.49\textwidth]{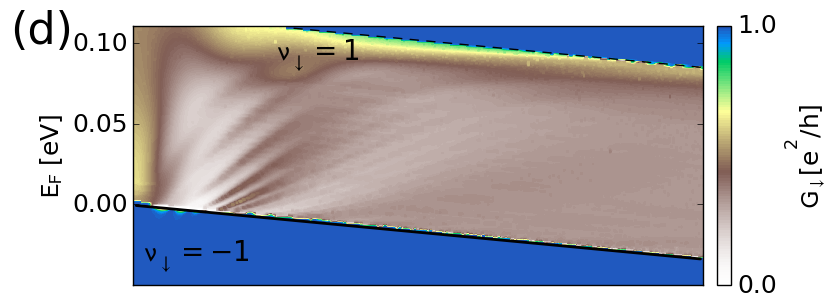}
 
 \includegraphics[width=0.49\textwidth]{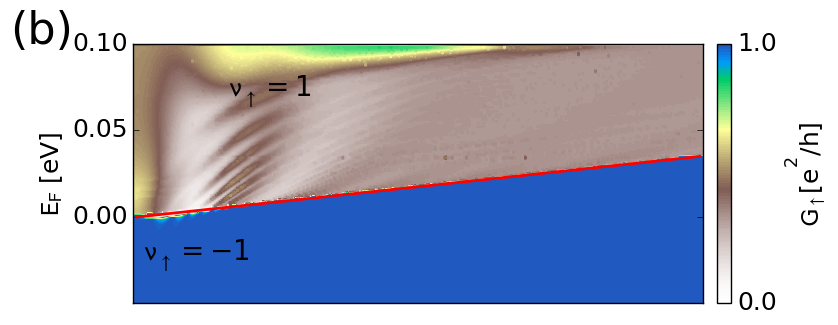}
 \includegraphics[width=0.49\textwidth]{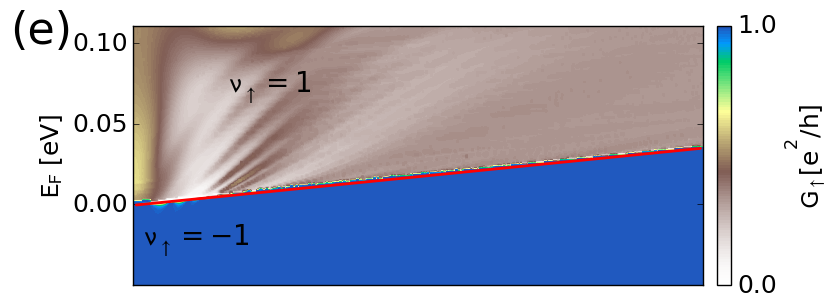}
 
 \includegraphics[width=0.49\textwidth]{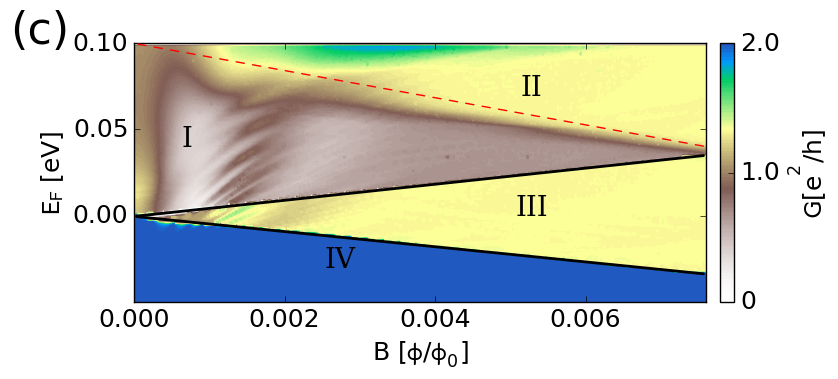}
 \includegraphics[width=0.49\textwidth]{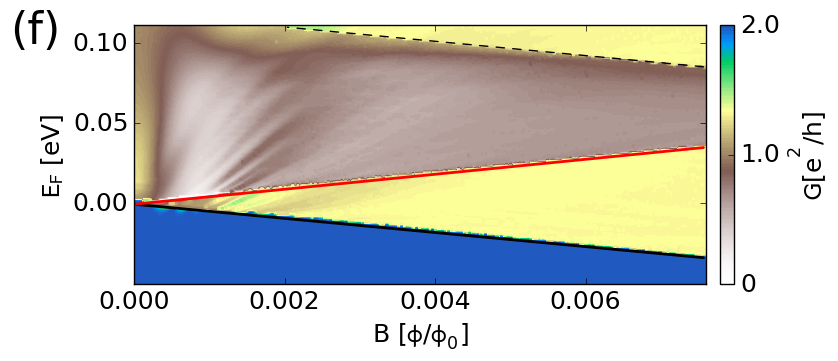}
  \caption{The conductance with dephasing by B\"uttiker probes in the metallic nanoribbon for (a,d) spin down, (b,e) spin up, (c,f) sum over both spin directions. The red (black) solid lines show where the number of spin up (down) channels changes. The red (black) thick solid  line show the Dirac point for spin up (down). The dashed lines show approximately where the junction states decouple from the edge states. In (c) the 
  Roman numbers label the regions of different equilibration regime described in the main text.
The hopping parameter between the B\"uttiker probes and the graphene layer is $t_B=0.5t$.
  } \label{inkoherentKr}
\end{figure*}

\begin{figure*}[tb!]
 \includegraphics[width=0.49\textwidth]{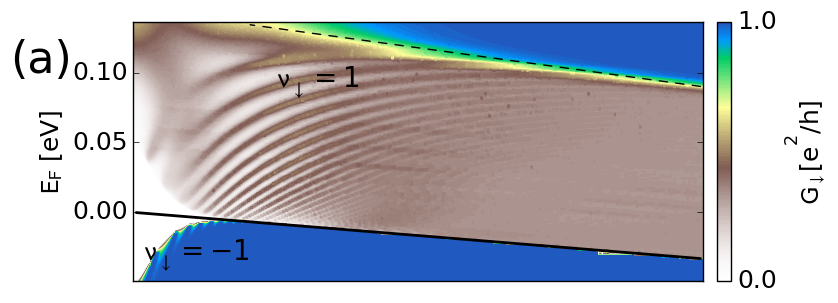}
 \includegraphics[width=0.49\textwidth]{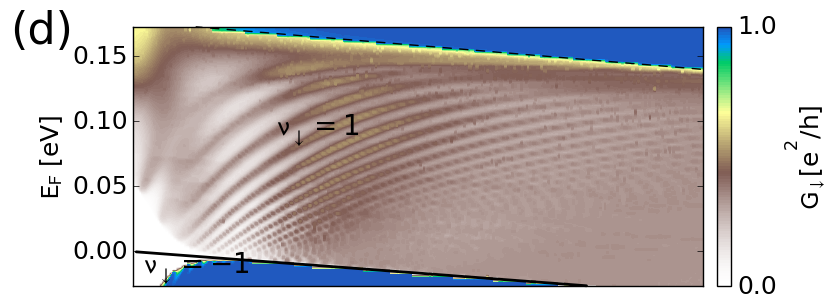}
 
 \includegraphics[width=0.49\textwidth]{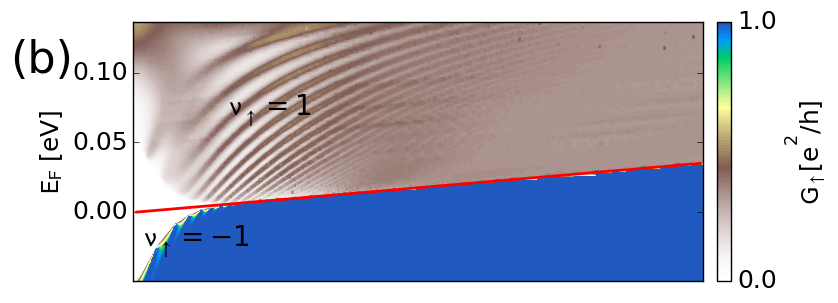}
 \includegraphics[width=0.49\textwidth]{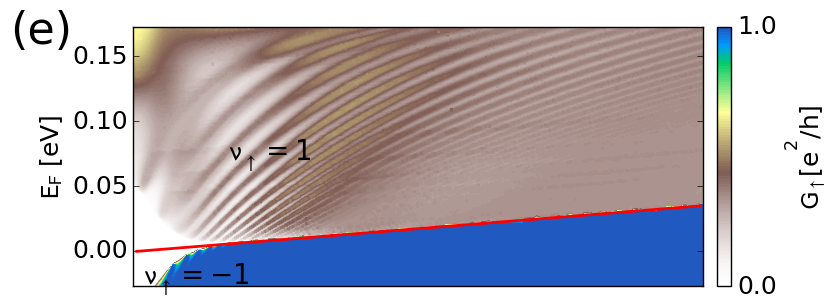}
 
 \includegraphics[width=0.49\textwidth]{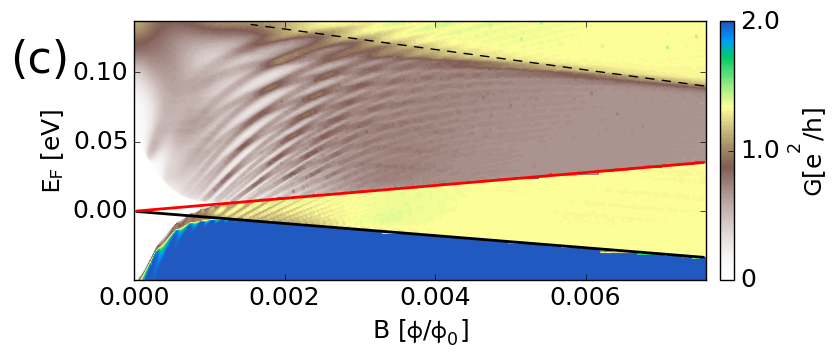}
 \includegraphics[width=0.49\textwidth]{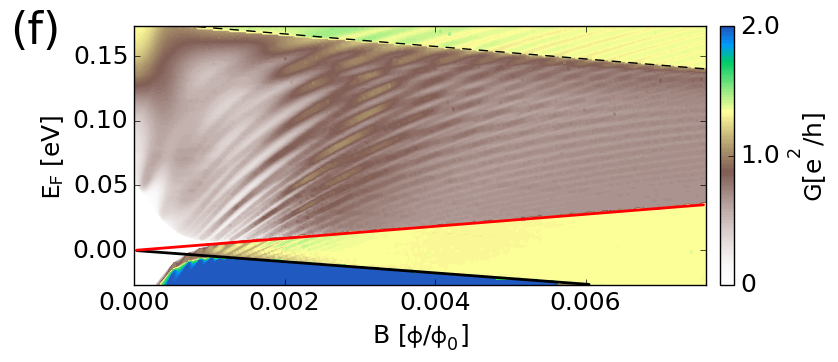}
  \caption{Same as in Fig.~\ref{inkoherentKr}, but for the semiconducting nanoribbon.
  } \label{inkoherentKrSem}
\end{figure*}

Figure~\ref{inkoherentKr} (Figure~\ref{inkoherentKrSem})  shows the conductance as a function of magnetic field and Fermi energy for the metallic (semiconducting) ribbon in presence of the equilibration.
In both cases the oscillations that were present for the fully coherent transport in Fig.~\ref{koherentKr} and Fig.~\ref{koherentKrSem} are suppressed at high magnetic field. In the limit of strong
dephasing the results for the semiconducting and metallic ribbon becomes nearly identical. 

\subsection{Discussion}

The results for equilibrated currents shown above involved electron and hole channels from the zeroth Landau level  along the bipolar junction.
 Whenever there is a bipolar junction in the considered nanoribbon, in the region under the tip and beyond there are only the channels from the lowest Landau level for magnetic field sufficiently high for the Hall effect to occur.
 For low Fermi energy when the junction diameter is bigger than the width of the ribbon, a nearly rectangular $n-p-n$ junction is induced. For such system with filling factors $+2/-2/+2 $ in the $n-p-n$ regions, respectively, the conductance with complete equilibration is predicted to be $\tfrac{2}{3} \tfrac{e^2}{h}$ \cite{Ozyilmaz2007}. In pure graphene
the spin-orbit coupling is negligible thus the mixing occurs only between same-spin channels  \citep{Amet2014, Weie2017}.

\begin{figure}[tb!]
 \includegraphics[width=0.8\columnwidth]{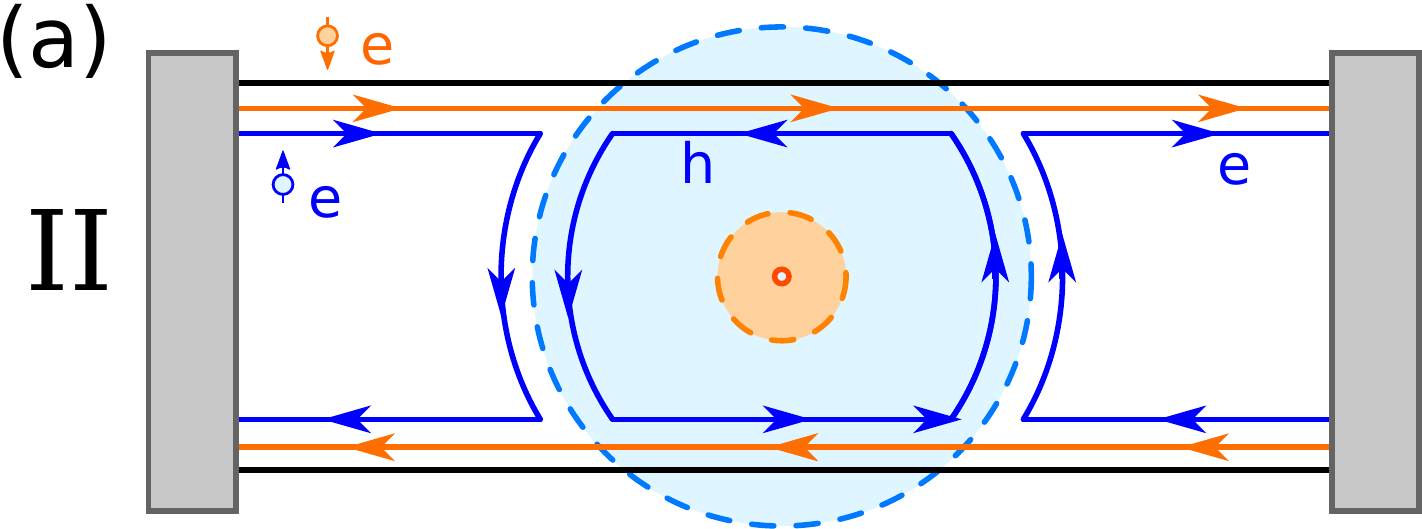}
 \includegraphics[width=0.8\columnwidth]{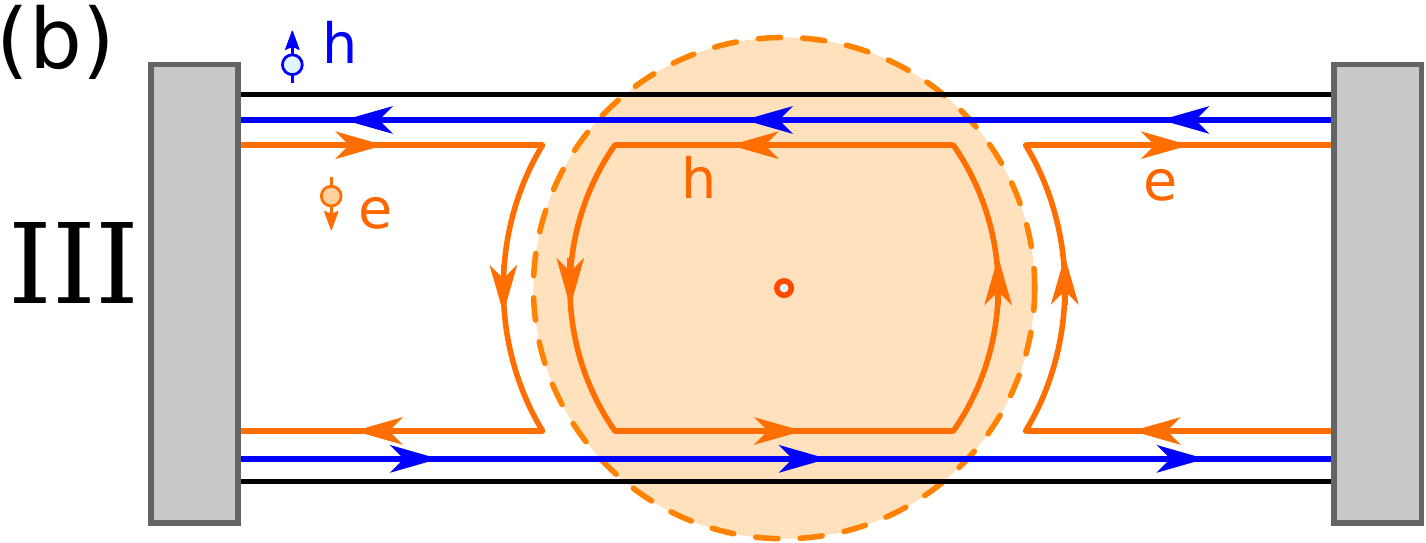}
  \caption{The schematic of the current propagation in the system with Zeeman split opposite spin channels corresponding to the area of the plot labeled in Fig.~\ref{inkoherentKr}(c) by Roman numbers II and III. (a) For spin up an $n-p-n$ junction is formed, while for spin down the circular junction is so small that the coupling between edge and junction states vanishes. The equilibration is only for spin up. (b) For spin down an $n-p-n$ junction is formed, while for spin up in the entire ribbon there is hole-like conductance. 
  The hole or electron like channels are labeled by ,,e'' or ,,h'', respectively, and the spin direction is color-coded, as shown by the tiny arrows with circles. The light-blue circle shows the extent of the tip-induced $n-p$ junction for the spin-up carriers, and the smaller orange circle for spin-down carriers. 
  } \label{schemeEquilib}
\end{figure}

In our study, we take into account the Zeeman splitting in the transport. 
In high magnetic field, the system can be tuned into configurations of filling factors with opposite spins, e.g. shown in Fig.~\ref{schemeEquilib}, and to conditions where the Fermi level is aligned
with the conduction band of one spin direction and with the valence band for the other. 
The coexistence of the electron and hole states close to neutrality point has been presented for graphene deposited on SiO$_2$ substrate \cite{Wiedmann2011, Abanin2007a}. 
However, samples on hexagonal boron nitride can be strongly insulating close to the neutrality point 
which was attributed by the valley symmetry breaking by the substrate \cite{Hunt2013, Amet2013}
with incommensurate lattice constant.


For the energy below the thick black line in Fig.~\ref{inkoherentKr}, in the region labeled by IV for both spins there is a single hole channel in the entire ribbon, and no bipolar junction occurs. The channels are transfered without backscattering and the conductance is $2 \tfrac{e^2}{h}$. In the region labeled by I for both spin orientation there is a single electron channel
outside the tip potential and a single-hole channel under the tip.
For a complete equilibration each of the spins should contribute $G_{\uparrow}=G_{\downarrow}=\tfrac{1}{3}\tfrac{e^2}{h}$ to the overall conductance. 
For $t_B=0.5t$ in region I of Fig.~\ref{inkoherentKr} the conductance still oscillates near $G=\tfrac{2}{3}\tfrac{e^2}{h}$. In the region III, for spin down carriers an $n-p$ junction forms, and for spin up only hole-conductance occurs as shown schematically in Fig.~\ref{schemeEquilib}(b). Thus with a complete equilibration, the spin down current contributes $G_{\downarrow}=\tfrac{1}{3}\tfrac{e^2}{h}$ and spin up  $G_{\uparrow}=1\tfrac{e^2}{h}$, and the summed conductance is $G=G_{\uparrow}+G_{\downarrow}=\tfrac{4}{3}\tfrac{e^2}{h}$. On the other hand, in the region labeled by II for spin up there is a broad $n-p$ junction. For spin down the Fermi energy is so high that either there is no junction or it has a very small radius so that the junction states are decoupled from the edge states, resulting in a lack of backscattering. Thus $G_{\uparrow}=\tfrac{1}{3} \tfrac{e^2}{h}$, $G_{\downarrow}=1 \tfrac{e^2}{h}$, and again $G=\tfrac{4}{3}\tfrac{e^2}{h}$. This situation is shown schematically in Fig.~\ref{schemeEquilib}(a). In the numerical calculations we obtain for regions II and III conductance close to $\tfrac{4}{3}\tfrac{e^2}{h}$ with only very small oscillations that vanish slowly in the high magnetc field limit. 

\section{Summary and Conclusions}

We studied an electron interferometer induced by potential of a floating gate within a graphene nanoribbon in the quantum Hall conditions. A numerical description of the transport conditions 
was given with the atomistic tight-binding Hamiltionian, tip potential derived from the electrostatics of the device and B\"uttiker virtual probes inducing dephasing effects at the junctions.
 The transport phenomena occurring in the interferometer involve the coherent Aharonov-Bohm oscillations and incoherent equilibration process resulting in appearance of fractional conductance plateaus. An interplay of the conductance oscillations and quantum Hall effects was explained.

We pointed out that the metallic ribbon is more effective in the screening of the top gate potential,
that the visibility of the coherent conductance oscillation is much larger for the semiconducting ribbon
and that in the strong dephasing limit of a complete equilibration the same results are obtained for the metallic and semiconducting ribbons with the fractional conductance plateaus replacing the current oscillations.

\section*{Acknowledgments}
This work was supported by the National Science Centre (NCN) according to decision DEC-2015/17/B/ST3/01161
and by AGH UST budget with the subsidy of the Ministry of
Science and Higher Education, Poland with Grant No. 15.11.220.718/6 for young researchers 
and Statutory Task No. 11.11.220.01/2.
The calculations were performed on PL-Grid Infrastructure. 

\section{Appendix: finite difference approach for the Poisson equation}
To obtain the entire potential in the structure, we solve the Poisson equation
\begin{equation}
 -\nabla \left (\nabla\epsilon(z) V\left(\mathbf{r} \right) \right) =4\pi\rho\left(\mathbf{r} \right),
 \label{Poisson1}
\end{equation}
where $\rho\left(\mathbf{r} \right)$ is the electron density originating from the spatial distribution of the electrons in graphene flake and it lies exactly at the interface between two media (see Fig. \ref{Struct1}(a)) i.e. for $z=0$.
The charge density is defined on the hexagonal graphene lattice. We effectively spread this quantity over the entire ribbon with finite thickness of $\Delta z$.
The $z$- dependent dielectric constant $\epsilon(z)$ is equal $\epsilon=3.9$ for $z<0$ (the Silicon Dioxide substrate) and $\epsilon=0$ for $z>0$ (vacuum).
In the regions where $|z|>\Delta z$ the equation \reff{Poisson1} simplifies to 
\begin{equation}
 -\epsilon \nabla^2 V\left( \mathbf{r} \right) =0,
\end{equation}
which is charge-free Laplace equation. 

We solve this equation numerically with finite difference method. We work on 3D cartesian grid with grid spacing equals to $ \Delta x=\Delta y=0.3$ nm and $ \Delta z=0.2$ nm and 
the total number of grid points $ 399\cdot 215\cdot 236\approx 2\cdot 10^7$.

\begin{figure}[htbp]
 a)\includegraphics[scale=0.17]{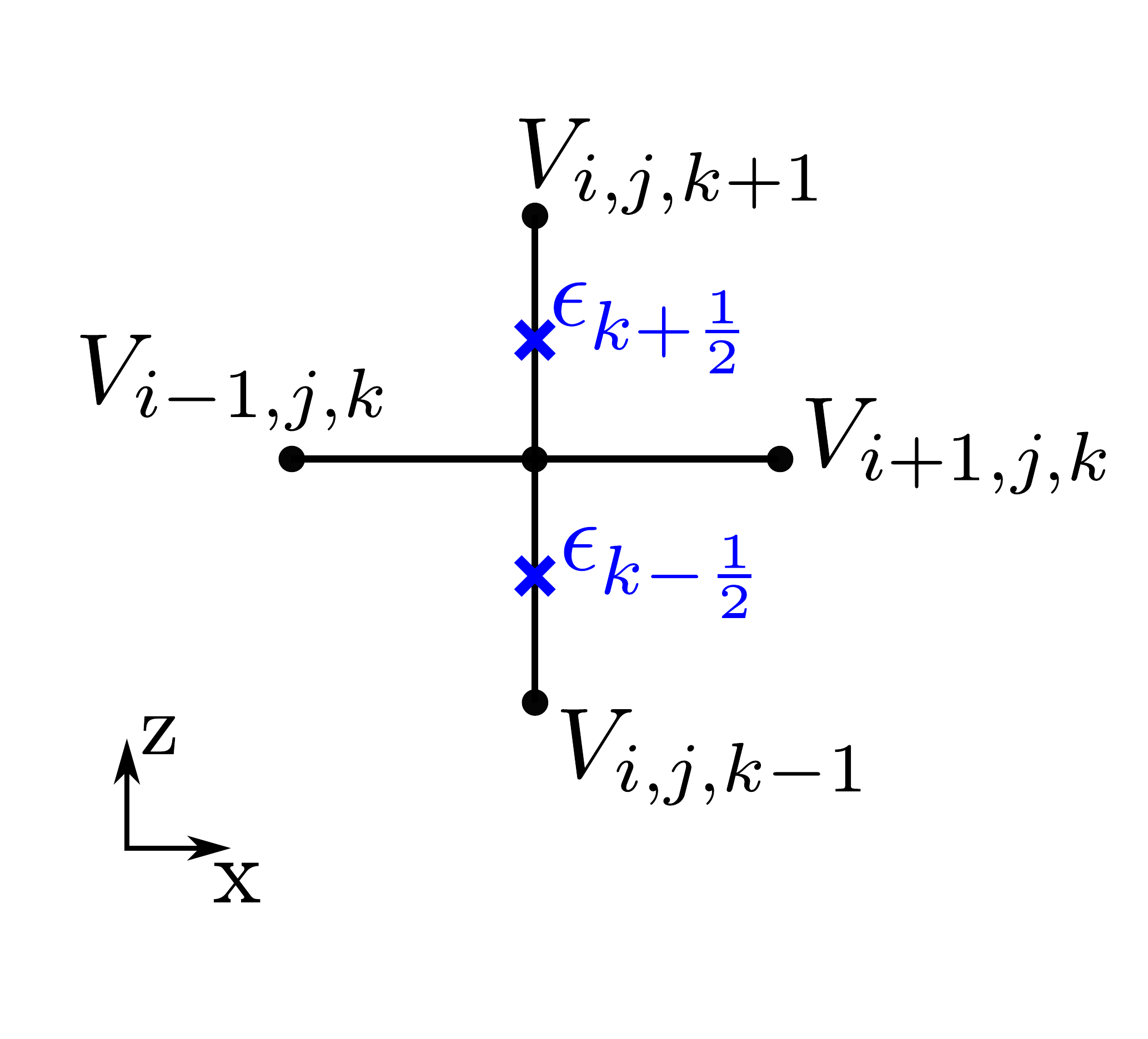}  b)\includegraphics[scale=0.17]{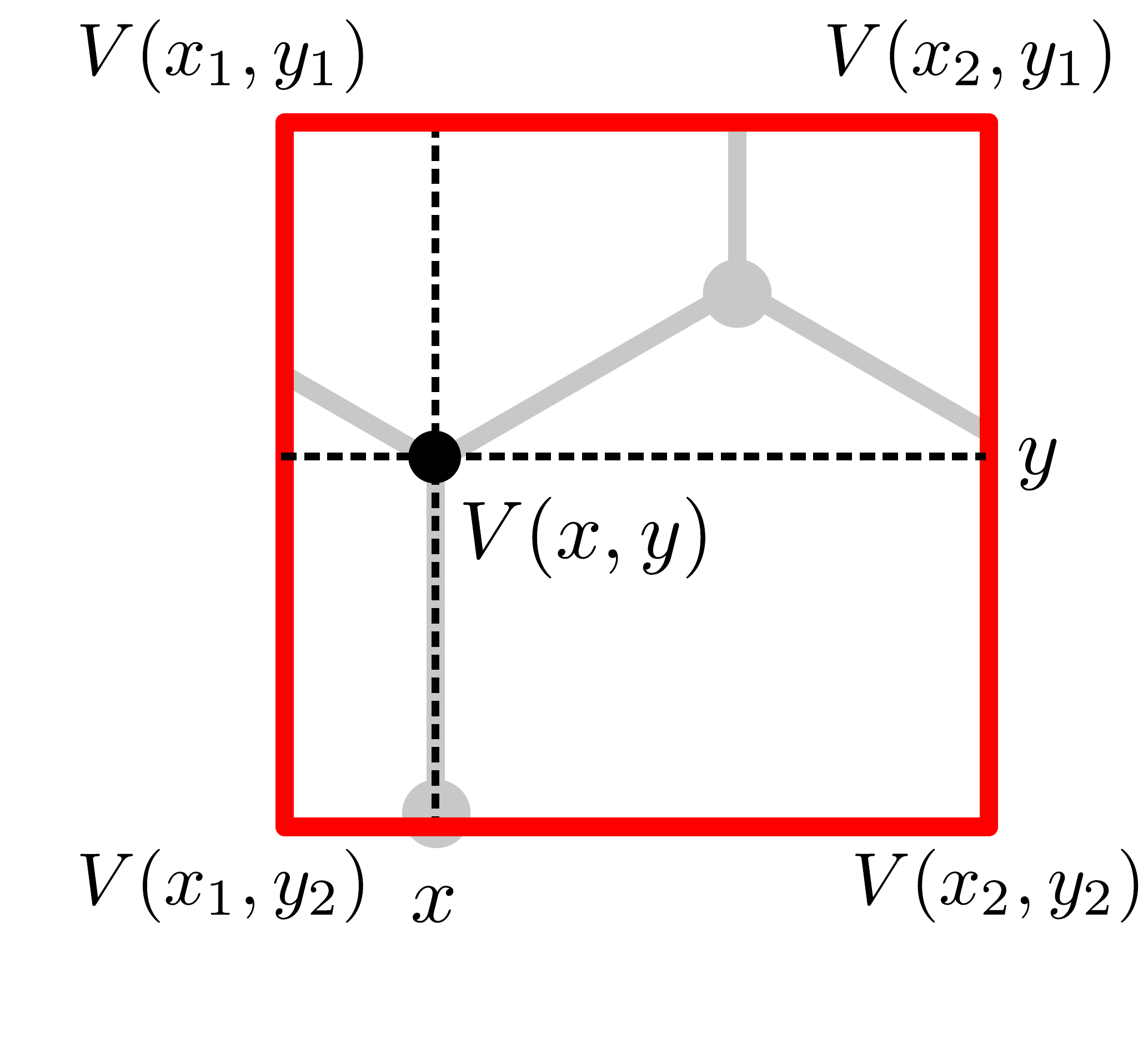}
 \caption{(a)The discretization stencil for the Poisson equation in the presence of abruptly changing dielectric constant. We define the dielectric constant
 at a mesh of points shifted in the $z$ direction. (b) Red square represents the Cartesian grid cell for $z=0$ used for Poisson equation calculations. The gray points and edges represent
 the graphene lattice. To obtain the potential value on the graphene lattice $(x,y)$ we interpolate its value using the four closest lattice points.}
 \label{PoissonDiscretImage1}
\end{figure}

For the charge-free equation \reff{PoissonDiscret1} we simply use the seven-point stencil (see Fig. \ref{PoissonDiscretImage1} (a)) that produces the set of equation in the form
\begin{equation} \label{PoissonDiscret1}
\begin{split}
  &\frac{V_{i+1,j,k}+V_{i-1,j,k}+V_{i,j+1,k}+V_{i,j-1,k}-4V_{i,j,k}}{\Delta x^2} \\
  &+\frac{V_{i,j,k+1}+V_{i,j,k-1}-2V_{i,j,k}}{\Delta z^2}=0,
\end{split}
\end{equation}
with $V_{i,j,k}=V(x_i,y_j,z_k)$, where tuple $(x_i,y_j,z_k)$ represents the point on the discretized 3D lattice.

In the region where $z=0$, one needs to take into the consideration a steep change of $  \epsilon$ and the presence of the 
right hand side of the equation (electron density distribution). As $\epsilon$ changes abruptly at the interface between media, we discretize it on the lattice shifted by half a step in the $z$ direction (see Fig. \ref{PoissonDiscretImage1}),
while potential $ V\left( \mathbf{r} \right) $ discretization grid stays the same. With these settings, we arrive at the system of equations of the form
\begin{equation} \label{PoissonDiscret2}
\begin{split}
  &\epsilon_k'  \frac{V_{i+1,j,k}+V_{i-1,j,k}+V_{i,j+1,k}+V_{i,j-1,k}-4V_{i,j,k}}{\Delta x^2} \\
  &+\frac{\epsilon_{k-\frac{1}{2} }V_{i,j,k-1}  +\epsilon_{k+\frac{1}{2}}V_{i,j,k+1}-2\epsilon_k'V_{i,j,k}}{\Delta z^2}   \\
  &=-4\pi\rho_{i,j,k},
\end{split}
\end{equation}
with $ \epsilon_k'$ defined as average between $z=0$ point i.e. $\epsilon'_k=\frac{1}{2}\left ( \epsilon_{k+\frac{1}{2}}+\epsilon_{k-\frac{1}{2}}    \right) $.

To solve equations \reff{PoissonDiscret1} and \reff{PoissonDiscret2}  we need to define the boundary conditions.
We consider Dirichlet and Neumann boundary conditions in this work. The first one accounts for the metallic gates: the bottom gate and the tip. This type of boundary
is incorporated into the grid equations by setting $V_{i_b,j_b,k_b}=V_g$, where tuple $i_b,j_b,k_b$ stands for the point on the boundary, and $V_g$ is the gate potential. The Neumann boundary is responsible for vanishing of
an electric field in the direction normal to the boundary, $ \frac{dV\left(\mathbf{r} \right)}{dn_B}=0$. To use this equation for a top wall, we discretize the condition in the $z$ direction obtaining
\begin{equation}
 \frac{V_{i,j,k+1}-V_{i,j,k-1}}{2 \Delta z}=0, \ \ V_{i,j,k+1}=V_{i,j,k-1},
\end{equation}
and plug this into the equations \reff{PoissonDiscret1}. For the side walls we use this condition in the same manner with a different direction of the normal vector $n_B$. 
To solve this set of equations we use the iterative successive over-relaxation method (SOR). For convergence, we check the $l^2$ norm of
the $V_{i,j,k}$ grid points and we stop if difference between iterations is less than $10^{-9}$ V.

After convergence is reached we have solution defined on the Cartesian grid. To obtain the value of the potential for the graphene lattice point $(x,y)$ (see Fig. \ref{PoissonDiscretImage1} (b)) we interpolate 
its value as $V(x,y)\approx a_0+a_1x+a_2y+a_3xy$, where the coefficients are the solution of the set of equations given by

\[
\begin{bmatrix}
1 & x_1 & y_1 & x_1y_1\\
1 & x_1 & y_2 & x_1y_2\\
1 & x_2 & y_1 & x_2y_1\\
1 & x_2 & y_2 & x_2y_2\\
\end{bmatrix}
\begin{bmatrix}
a_0\\
a_1\\
a_2\\
a_3
\end{bmatrix}=
\begin{bmatrix}
V(x_1,y_1)\\
V(x_1,y_2)\\
V(x_2,y_1)\\
V(x_2,y_2)
\end{bmatrix}.
\]

%

\bibliographystyle{apsrev4-1}
\bibliography{tip}

\end{document}